\documentclass[twoside]{article}
\usepackage[accepted]{aistats2020}
\usepackage{enumerate}
\usepackage[round]{natbib}

\usepackage[utf8]{inputenc} % allow utf-8 input
\usepackage[T1]{fontenc}    % use 8-bit T1 fonts
\usepackage{url}            % simple URL typesetting
\usepackage{booktabs}       % professional-quality tables
\usepackage{amsfonts}       % blackboard math symbols
\usepackage{nicefrac}       % compact symbols for 1/2, etc.
\usepackage{microtype}      % microtypography

\usepackage{paralist}
\usepackage{microtype}
\usepackage{graphicx}
\usepackage{subfigure}
\usepackage{booktabs} % for professional tables
\usepackage{amsthm,soul}
\usepackage{amsmath, amssymb}
\usepackage{mathtools}
\usepackage{enumitem}
\usepackage{cuted}
\usepackage{graphics}
\usepackage{wrapfig}
\usepackage{color}
\definecolor{hcitecolor}{RGB}{40,180,40}
\usepackage[colorlinks=true,citecolor=blue, linkcolor=blue]{hyperref}
\newtheorem{theorem}{Theorem}
\newtheorem*{theorem*}{Theorem}
\newtheorem{assumption}{Assumption}
\newtheorem*{assumption*}{Assumption}

\newtheorem*{corollary*}{Corollary}

\newtheorem*{lemma*}{Lemma}

\newtheorem*{remark*}{Remark}

\newcommand{\ccmXY}{\mathrm{CCM}(X_t \mid Y_t)}
\newcommand{\ccmYX}{\mathrm{CCM}(Y_t \mid X_t)}

\newcommand{\Y}{\widetilde{Y}}
\newcommand{\X}{\widetilde{X}}

\newcommand{\CCMSCM}{CCM+SCM}

\begin{document}
\runningtitle{Dynamical Systems Theory for Causal Inference with Application to Synthetic Control Methods}


\twocolumn[

\aistatstitle{Dynamical Systems Theory for Causal Inference \\ with Application to Synthetic Control Methods}

\aistatsauthor{Yi Ding \And Panos Toulis}

\aistatsaddress{University of Chicago\\Department of Computer Science \And  
University of Chicago \\ Booth School of Business} ]

\begin{abstract}
In this paper, we adopt results in nonlinear time series analysis for causal inference in dynamical settings.~Our motivation is policy analysis with panel data, particularly 
through the use of ``synthetic control'' methods.
These methods regress pre-intervention outcomes of the treated unit to outcomes from a pool of control units, and then use the fitted regression model to estimate causal effects post-intervention.
In this setting, we propose to screen out control units that have a weak dynamical relationship to the treated unit.
% according to well-established measures of relationship strength in dynamical systems theory.
%
In simulations, we show that this method can mitigate bias from 
``cherry-picking'' of control units, which is usually an important concern.
We illustrate on real-world applications, including the tobacco legislation example of \citet{Abadie2010}, and Brexit.
\end{abstract}

\section{Introduction} \label{sec:intro}
In causal inference, we compare outcomes of units who received the treatment with outcomes from units who did not. A key assumption, often made implicitly, is that 
the relationships of interest are static and invariant. For example, in studying the effects of schooling on later earnings, we usually consider potential outcomes~$Y_i(k)$, for some student $i$ had the student received $k$ years of schooling. Since only one potential outcome can be observed for each student, causal inference needs to rely on comparisons between students who received varying years of schooling. The validity of the 
results therefore rests upon the assumption that the relationship between years of schooling and earnings is temporally static and unidirectional. 

\begin{figure}[t!]
	\centering
	\includegraphics[width=0.4\textwidth]{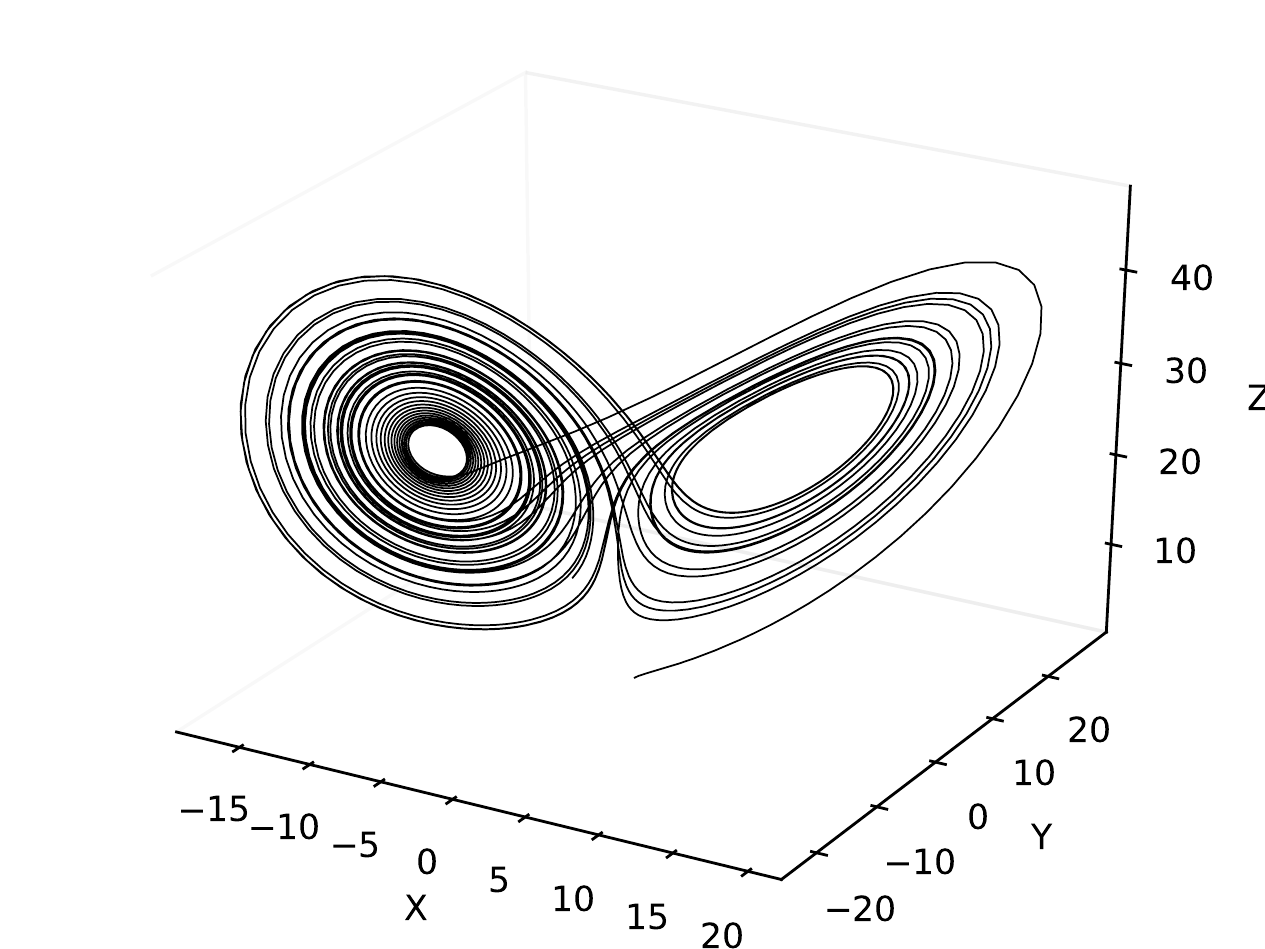}
	\vspace{-0.1in}\caption{A Lorenz attractor plotted in 3D.}\label{fig:lorenz}   \vspace{-0.15in}
\end{figure}

However, in many real-world settings, different variables exhibit dynamic interdependence, sometimes showing positive correlation and sometimes negative. Such ephemeral correlations can be illustrated with a popular dynamical  system shown in Figure~\ref{fig:lorenz}, the Lorenz system~\citep{Lorenz1963}. The trajectory resembles a butterfly shape indicating varying correlations at different times: in one wing of the shape, variables $X$ and $Y$ appear to be positively correlated, and in the other they are negatively correlated. Such dynamics present new methodological challenges for causal inference that have not been addressed. In relation to the  schooling example, our analysis of schooling effect on earnings could occur on one wing of the system, where the correlation is, say, positive. However, crucial policy decisions, such as college subsidies, could occur on the other wing where the relationship is reversed. Such discord between data analysis and policy making is clearly detrimental to policy effectiveness.

Despite its longstanding importance in many scientific fields, dynamical systems theory has not 
found a way into modern causal inference~\citep{Durlauf2005}. The main goal of this paper is to leverage key results from dynamical systems to guide causal inference in the presence of dynamics. For concreteness, we focus on synthetic control methods~\citep{Abadie2010}, which are popular for 
policy analysis with panel data. 
Our methods, however, can more generally be applied when causal  inference involves some form of matching between treated and control units in the time domain.

\section{Preliminaries}
\label{sec:prelims}
Here, we give a brief overview of comparative case studies with panel data to fix concepts and notation. 
Later, in Section~\ref{sec:method}, we describe our method.

Following standard notation, we consider $J+1$ units, with only one unit being treated. Let $Y_{it}^N$ be the potential outcome for unit $i$ at time $t$ in a hypothetical world where the intervention did not occur~(denoted by the exponent ``N''), where $i=1,2,\ldots,J+1$, and $t=1,2,\ldots,T$; 
also let $Y_{it}^I$ be the corresponding potential outcome assuming the intervention did occur.
Let $D_{it}$ be a binary indicator of whether unit $i$ is treated at time $t$. By convention, 
and without loss of generality,
only unit 1 receives the treatment, and there exists $T_0\in(0, T)$  such that
$$
D_{it} = 1,~\text{if and only if}~i=1~\text{and}~t > T_0.
$$
%
%Assuming $i=1$ as the treated unit is a convention and is without loss of generality.
The observed outcome for unit $i$ at time $t$, 
denoted by $Y_{it}$, therefore satisfies:
\begin{align}
Y_{it}=Y_{it}^N + (Y_{it}^I-Y_{it}^N) D_{it},
\end{align}
where $\tau_{it} = Y_{it}^I-Y_{it}^N$ is the causal effect of intervention on unit $i$ at time $t$.
Suppose there exist weights $w_2, \ldots, w_{J+1}$ such that 
$\sum_{j=2}^{J+1} w_j = 1$ and $w_j \ge 0$, and
$Y_{1t}^N = \sum_{j=2}^{J+1}  w_j  Y_{jt}$, 
for $t=1, \ldots, T_0$. Then, the causal effect of the intervention can be estimated through:
\begin{align}
\label{eq:syn}
\hat\tau_{1t} = Y_{1t} - \sum_{j=2}^{J+1} w_j Y_{jt},~\text{for every}~t>T_0.
\end{align}
The time series defined with the term $\sum_{j=2}^{J+1} w_j Y_{jt}$ in Equation~\eqref{eq:syn} is the {\em synthetic control}. 
This synthetic control unit can be construed to be representative of the treated unit~($i=1$)
had the treated unit not received treatment.
Because of the constraints put on $w_j$, namely that they are nonnegative and sum to one, the fitted values of the weights reside on the edges of a polytope, and so many weights are set to 0. Such sparsity in the weights corresponds to control selection, and so only a few control units are used to model the outcomes of the treated unit.

The synthetic control methodology is an important example of comparative case studies~\citep{Angrist2008, Card1994}, and generalizes other well-known methods, such as ``difference-in-differences''. As a methodology it is  simple and transparent, and so synthetic controls have become widely popular in the fields of policy analysis~\citep{Abadie2010,Kreif2016, shaikh2019randomization}, criminology~\citep{Saunders2015}, politics~\citep{Abadie2003,Abadie2015}, and economics~\citep{Billmeier2013}.

Theoretically, the treatment effect estimator, $\hat\tau_{1t}$, is asymptotically unbiased as the number of pre-intervention periods grows when the outcome model is linear in (possibly unobserved) factors and the treated unit ``lives'' in the convex hull of the controls~\citep[Theorem 1]{Abadie2010}. As such, a key assumption of model continuity is implicitly made for identification, where the weights $w_j$ are assumed to be time-invariant. Furthermore, control selection in synthetic controls depends only on the statistical fit between treated and control outcomes in the pre-intervention period, which opens up the possibility of cherry-picking controls to bias causal inference. 
In the following section, we illustrate these problems with an example.

\subsection{Motivation: an Adversarial Attack to the Synthetic Control Method}
\label{sec:motivation}

As a motivation, we use the example of California's tobacco control program in 1989, as described in the original paper of synthetic controls~\citep{Abadie2010}.
The goal is to estimate the effect of Proposition 99, a large-scale tobacco control program passed by electorate majority in 1988 in California.
The proposition took effect in 1989 through a sizeable tax hike per cigarette packet. The panel data include annual state per-capita cigarette sales from 1970 to 2000 as outcome, along with related predictors, such as state median income and \%youth population. We have a pool with 38 states as potential controls, after discarding states that adopted similar programs during the 1980's.

The synthetic control methodology proceeds by calculating a weighted combination of control unit outcomes to fit cigarette sales of California, using only pre-1989 data. In this application, the weighted combination is: Colorado (0.164), Connecticut (0.069), Montana (0.199), Nevada (0.234), and Utah (0.334), where the numbers in the parentheses are the corresponding weights. 
The implied model is the following:
\begin{align}
\label{eq:ca_example1}
%\resizebox{.9\hsize}{!}{$\widehat{Y}_{\mathrm{CA}, t} = 0.164 \times Y_{\mathrm{CO}, t} + 0.069 \times Y_{\mathrm{CT}, t}
%	+ 0.199 \times Y_{\mathrm{MO}, t} 
%	+ 0.234\times Y_{\mathrm{NV}, t} +
\widehat{Y}_{\mathrm{CA}, t} &= 0.164 \times Y_{\mathrm{CO}, t} + 0.069 \times Y_{\mathrm{CT}, t} \\ \nonumber
	&+ 0.199 \times Y_{\mathrm{MO}, t} 
	+ 0.234\times Y_{\mathrm{NV}, t} +
	0.334 \times Y_{\mathrm{UT}, t},
\end{align}
where $Y$ denotes packet sales at a particular state and time (a state is denoted by a two-letter code; e.g., $\mathrm{CA}$ stands for California). 
We note that time $t$ in the model of Equation~\eqref{eq:ca_example1} is before intervention ($t \leq 1989$), 
so that all states in the data, including California, are in control for the entire period considered in the model.

\begin{figure*}[t!]
	\centering
	\begin{minipage}[b]{0.4\linewidth}
		\includegraphics[width=1\textwidth]{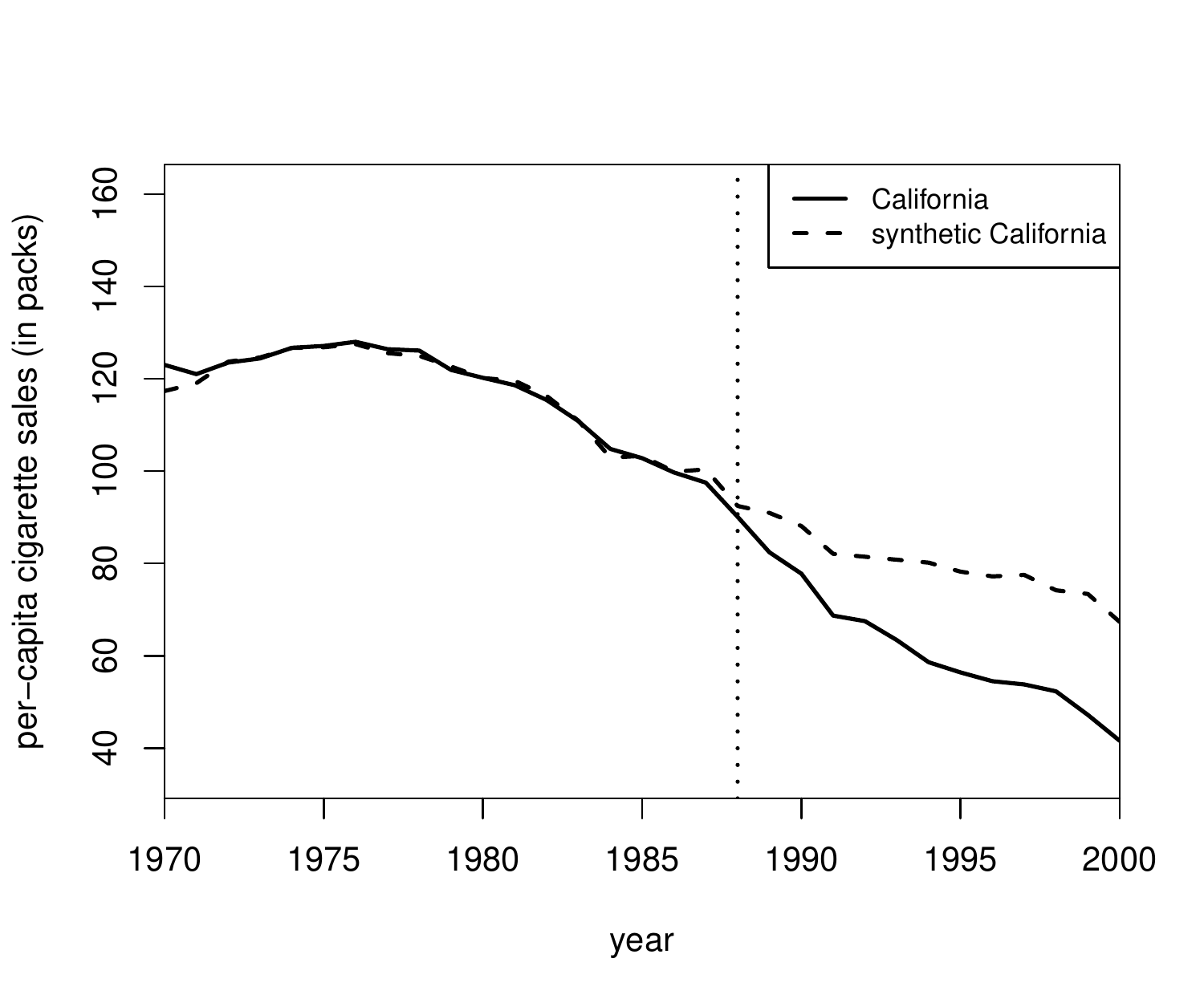}  
	\end{minipage}
	\begin{minipage}[b]{0.4\linewidth}
		\includegraphics[width=1\textwidth]{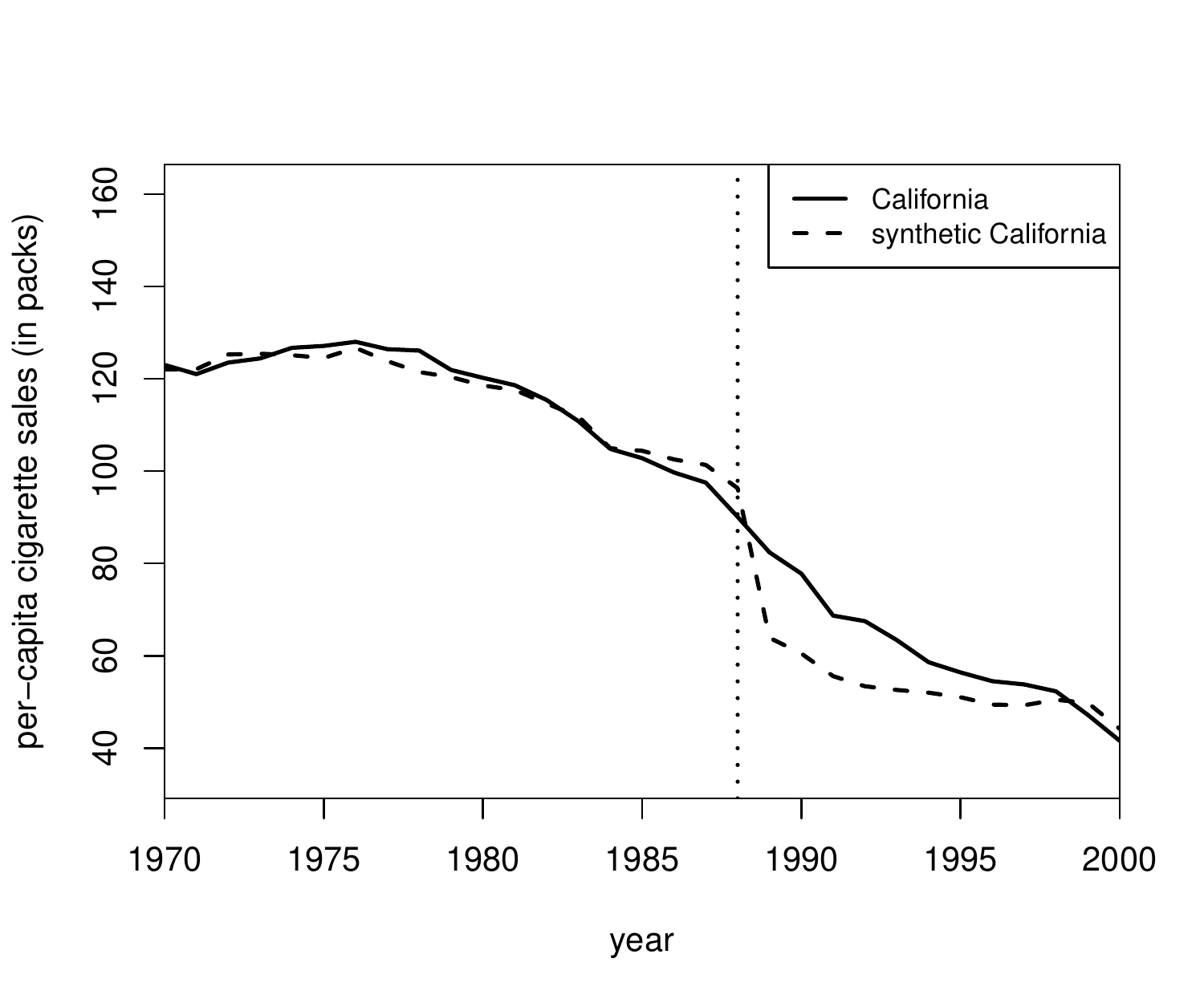}
	\end{minipage}
	\vspace{-10pt}
	\caption{Trends in per-capita cigarette sales. The solid line is actual California and the dashed line is synthetic California, while the vertical line indicates time of intervention. \textbf{Left}: original setting. \textbf{Right}: adversarial setting where the synthetic California is created according to Equation~\eqref{eq:ca_example2}.}\label{fig1}  
\end{figure*}

The idea for causal inference through this approach is that the same model in Equation~\eqref{eq:ca_example1} can be used to estimate the counterfactual outcomes for California, $Y_{\mathrm{CA}, t}$, for $t > 1989$, had California not been treated with the tax hike in 1989. 
By comparing the post-intervention  data from actual California that 
was  treated with the tobacco control program in 1989, and predictions for {\em synthetic California} that hypothetically stayed in control in 1989,  we can estimate that per-capita cigarette sales reduced by 19 packs on average by Proposition 99, suggesting a positive causal effect. This is illustrated in the left figure of Figure~\ref{fig1}.
As mentioned earlier, an implicit assumption here is that of model continuity:  we assume that the same model that fits pre-intervention California can be used to predict the counterfactual outcomes of a post-intervention, non-treated California.

This model continuity assumption relies critically on the choice of control units in the model of Equation~\eqref{eq:ca_example1}. Currently, this choice relies mostly on the subject-matter expert, which leaves open opportunities for cherry-picking in constructing the control pool. To illustrate this problem we can perform the following manipulation.
First, we add 9 unemployment-related time series\footnote{Data from the Local Area Unemployment Statistics (LAUS) program of the Bureau of Labor Statistics~\citep{BLS2018}. See Supplement for details.}, namely $Y'_{\mathrm{AD1},t},\ldots,Y'_{\mathrm{AD9},t}$, into the pool of potential controls, where ``$\mathrm{AD}$'' stands for ``adversarial''. Second, before adding these units to the control pool we transform the time series as follows:
$Y_{\mathrm{AD}i,t} = Y'_{\mathrm{AD}i,t} -50 + 90\cdot\mathbb{I}\{t \le 1989\}$. 
This transformation ensures that the adversarial time series has similar scale 
to time series on cigarette consumption before treatment.
Since the synthetic control method only relies on statistical fit, it may pick up the artificial time series from the new control pool. Indeed, the new synthetic California 
is now described by the following model:
\begin{align}
\label{eq:ca_example2}
\widehat{Y}_{\mathrm{CA}, t} &= 0.247 \times Y_{\mathrm{CO}, t} + 0.179 \times Y_{\mathrm{CT}, t} \\\nonumber 
&+ 0.196\times Y_{\mathrm{NV}, t} + 0.06 \times Y_{\mathrm{NH}, t} 
+ 0.011 \times Y_{\mathrm{WY}, t} \\\nonumber 
&+ 0.3 \times Y_{\mathrm{AD}\ast, t},
\end{align}
where $Y_{\mathrm{AD}\ast, t}$ represents that the adversarial time series was selected --- the specific index is irrelevant. This produces a new synthetic California that is drastically different than before. The weight on the artificial control unit is in fact the highest compared to the weights on all other units, which is clearly undesirable.
More importantly, with the new synthetic control California, we estimate a {\em negative} causal effect of 8 packs on average~(see right sub-figure in Figure~\ref{fig1}).

To address this problem, our paper leverages fundamental results in dynamical systems theory, 
such as time-delay embeddings.
The goal is to pre-screen control units based on how strongly related they are to treated units from a dynamical point of view. 
% Roughly speaking, the strength of relationship is  estimated by comparing the geometry of the aforementioned embeddings between treated and control units. 
The key idea is that state cigarette consumption data evolve on the same attractor, 
whereas adversarial time series do not. Thus, the latter should be less dynamically related to the treated state, 
and so they should be removed from the control pool. 

We note that our proposed method differs from recent work in synthetic controls, which 
has mainly focused on high-dimensional, matrix completion, or de-biasing methods~\citep{amjad2018robust,ben2018augmented,athey2018matrix, hazlett2018trajectory}.
These methods take a regression model-based approach, whereas we treat panel data 
as a nonlinear dynamical system. 
% Another related method is trajectory balancing~\citep{hazlett2018trajectory}, which replaces the linear projection of the treated unit to controls with a non-linear one using kernels. Although this method does not consider dynamical information as we do here, That said, a method ?CCM+TB? is conceivable, where CCM filtering happens on the first step, and trajectory balancing on the second.
More broadly, our approach shows that dynamical systems theory can be integrated into statistical frameworks of causal inference, a goal that so far has remained elusive~\citep{Rosser1999,Durlauf2005}.

\section{Methods}
\label{sec:method}
In this section, we describe our proposed method. In Section~\ref{sec:ccm}, we present the method of convergent cross mapping~(CCM), which is the fundamental building block of our method.
In Section~\ref{sec:theory}, we motivate CCM through a theoretical analysis on a 
simple, non-trivial time-series model. Finally, in Section~\ref{sec:procedure}, we
give details on our proposed method.

\subsection{Convergent Cross Mapping (CCM)}
\label{sec:ccm}
The basis of our approach is to consider the available panel data as a dynamical system.
In particular,  the state of the system at time $t$ is the collection of all time series, $(Y_{1t}, Y_{2t},\ldots,Y_{(J+1)t})$, where $J$ is the number of controls. Taken across all possible $t$, this implies a manifold, known as the {\em phase space}, denoted by 
$M=\left\{(Y_{jt} : j\in1, \ldots, J+1) : t\in[0, T]\right\}$, where $T$ denotes the length of time series, 
and is fixed. For example, when $J=1$ there are two units in total and $M$ is a curve (possibly self-intersecting) on the plane. 

A seminal result in nonlinear dynamics is Takens' theorem~\citep{Takens1981}, which shows that the phase space of a dynamical system can be reconstructed through time-delayed observations from the system.
Specifically, let us define a delay-coordinate embedding of the form
\begin{equation}
\label{eq:manifold}
\Y_{jt}=[Y_{jt}, Y_{j(t-\tau)},\ldots, Y_{j(t-(d-1)\tau)}],
\end{equation}
where $\tau>0$ is the time delay. The key theoretical result of~\cite{Takens1981} is that the manifold, $\widetilde M_j$, defined from outcomes $\{\Y_{jt}\}$ is diffeomorphic (i.e., the mapping is differentiable, invertible, and smooth) to the original manifold $M$, meaning that some important topological information is preserved, such as invariance to coordinate changes. In other words, $\widetilde M_j$ is a reconstruction of $M$. It follows that different reconstructions $\widetilde M_j$,  for various $j$, are diffeomorphic to each other, including the original manifold $M$, which implies cross-predictability. For two different reconstructions $\widetilde M_j$ and $\widetilde M_{j'}$, with their corresponding base time series $Y_{jt}$ and $Y_{j't}$,  we
could use $\widetilde M_j$ to predict $Y_{j't}$ and use 
$\widetilde M_{j'}$ to predict $Y_j$. By measuring this cross-predictability, the relative strength of dynamical relationship between any two variables in the system can be quantified~\citep{Schiff1996,Arnhold1999}.

One recent method utilizing this idea is convergent cross mapping~\citep[CCM]{Sugihara2012}. In addition to the idea of cross-predictability, CCM also relies on a smoothness implication of Takens' theorem, whereby neighboring points in the reconstructed manifold are close to neighboring points in the original manifold. This suggests that cross predictability will increase and stabilize as the number of data points grow. The cross predictability is quantified by a CCM score, which we will address later. 

Operationally, the generic CCM algorithm considers two time series, say $X_{t}$ and $Y_t$, and their corresponding delay-coordinate embedding vectors at time $t$, namely 
\begin{align}
\label{eq:embed}
&\X_t = [X_{t}, X_{(t-\tau)},\ldots, X_{(t- d\tau + \tau)}], \\ \nonumber
&\Y_t = [Y_{t}, Y_{(t-\tau)},\ldots, Y_{(t- d\tau + \tau)}],
\end{align}
where $d$ is known as the embedding dimension and 
$t\in\{1+(d-1)\tau, 1 + d\tau, \ldots, T\}$.
The manifold based on the phase space of $\Y_t$ is denoted by $M_Y$, and the manifold based on $\X_t$ is denoted by $M_X$, where the manifold definitions follow from Equation~\eqref{eq:manifold}.
The idea is that these manifolds are diffeomorphic to the original manifold of the dynamical system of $Y$ and $X$. A manifold from the delay embedding of one variable can be used to predict the other variable, and the quality of this prediction is an indication of which variable ``drives'' the other.

Such prediction proceeds in discrete steps as follows. First, we build a nonparametric model of $X_t$ using the reconstruction manifold based on $Y_{t}$. For a given time point $t$ we pick the $(d+1)$-nearest neighbors from $t\in\{1+(d-1)\tau, 1 + d\tau, \ldots, L\}$ in $\Y_{t}$, where $L<T$ is called the library size, and denote their time indices (from closest to farthest) as $\{t_1,\cdots,t_{d+1}\}$. A linear model for $X_t$ is as follows:
\begin{align}
\label{eq:ccm}
\widehat{X}_{t} = \sum\nolimits_{i=1}^{d+1} w(t_i,t) X_{t_i},
\end{align}
where $w(t_i,t)$ is the weight based on the Euclidean distance between $\Y_{t}$ and its $i$-th nearest neighbor on $\Y_{t_i}$, for example,
$w(t_i,t) = \exp(-d_i/d_1) / \sum_j \exp(-d_j/d_1)$ and $d_i = ||\Y_{t} - \Y_{t_i}||$, with the usual $L_2$ norm. The difference defined by mean absolute error (MAE) between $X_{t}$ and $\widehat{X}_{t}$ across $t$ is defined as the CCM score of $Y_{t}$ on $X_{t}$ (lower is better):
%\begin{align}
%	\label{def:ccm}
%	\mathrm{CCM}(X_{t} \mid Y_{t}) = \rho(X_{t}, \widehat{X}_{t}).
%\end{align}
\begin{align}
\label{eq:score}
&\ccmXY = |X_t - \widehat X_t|, \\ \nonumber
&\ccmYX = |Y_t - \widehat Y_t|. 
\end{align}
%In this definition, the time series are assumed to be normalized.

Intuitively, the CCM score captures how much information is in $Y_{t}$ about $X_{t}$.
For instance, if $X_{t}$ dynamically drives $Y_{t}$ we expect $\widehat{X}_{t}$ to be close to $X_{t}$. Similarly, $\mathrm{CCM}(Y_{t} \mid X_{t})$ is obtained by repeating the above procedure symmetrically, using $M_X$ of the values from the delay embedding $\X_t$ of $X_t$.
The value of $\mathrm{CCM}(Y_{t} \mid X_{t})$ quantifies the information in $X_{t}$ about $Y_{t}$. The two CCM scores jointly quantify the dynamic coupling between the two variables.
 As mentioned earlier, Takens' theorem implies that there exists a one-to-one mapping such that the nearest neighbors of $Y_{t}$ identify the corresponding time indices of nearest neighbors of $X_{t}$, if $X_{t}$ and $Y_{t}$ are dynamically related. As the library size, $L$, increases, the reconstruction manifolds $ M_X$ and $M_Y$ become denser and the distances between the nearest neighbors shrink, 
 and so the CCM scores will converge; see~\cite{Sugihara1990, Casdagli1991} for more details. 

From a statistical perspective, the CCM method is a form of nonparametric time-series estimation~\citep{hardle1997review}. The unique feature of CCM, and more generally of delay embedding methods, is that the nonparametric components are in fact the time indices $t_1, \ldots, t_{d+1}$ in Equation~\eqref{eq:ccm}. This differs from, say, kernel smoothing~\citep{Hastie2001}, where the target point $X_{t}$ is fitted by neighboring  observations to smooth estimation.
Importantly, CCM is not in competition with Granger causality~\citep{Granger1969}, but rather complements it. The key problem with Granger causality is that it requires ``separability'' of the effects from different causal factors. This condition generally does not hold in real-world dynamical systems that exhibit so called ``weak coupling''. CCM is unique because it can work in such systems~\citep{Sugihara2012}. 
Finally, CCM is backed up by a growing literature in the physical sciences as it is tailored to dynamical complex systems~\citep{runge2019inferring}.

%From a practical perspective, CCM has been widely used with notable empirical success in ecology and climate science, in early warning signs of critical transitions in complex ecosystems, and in neuroscience, to name a few~\citep{Sugihara2012,YeE1569, heskamp2014convergent, tsonis2018convergent}.

\subsection{Theory of CCM on Autoregressive Model}
\label{sec:theory}
In this section, we illustrate CCM through an AR(1) autoregression model.
Of course, AR(1) is a simple model that most certainly does not capture the details of real-world time series. However, its simplicity allows us to do two things. 
First, we derive analytic formulas for the  CCM scores in Equation~\eqref{eq:score}. 
Due to CCM's nonlinear nature, such formulas are not easily attainable, in general. In fact, 
we are unaware of any other analytic expressions for CCM in the literature, so our work here makes a broader contribution. 
Second, we can compare the CCM formulas with the parameters of the AR(1) model to 
better understand CCM as a tool; this is only possible because AR(1) is simple enough that the strength of causal relationships between variables is discernible from the model parameters.

The outcome model we consider is as follows:
\begin{align}
\label{eq:AR}
X_t  = \alpha X_{t-1} + \mu + \epsilon_t,  \quad \quad
Y_t  = \beta X_{t-1} + \mu + \zeta_t,
\end{align}
where $\alpha$, $\beta$ are fixed, with $|\alpha|<1$, $\beta\geq 0$, $\mu$ is the drift, $\epsilon_t \sim N(0, \sigma_X^2)$ and $\zeta_t \sim N(0, \sigma_Y^2)$ are zero-mean and constant-variance normal errors, with $\sigma_Y > \sigma_X$.
In this joint dynamical system of $X$ and $Y$, it is 
evident that $X$ generally drives $Y$ since $X$ evolves independently of $Y$, whereas the evolution of $Y$ depends on $X$. We are interested in knowing how CCM quantifies this asymmetric dynamic relationship between $X$ and $Y$, 
and whether it captures the dependence on parameters $\alpha, \beta, \mu, \sigma_X^2, \sigma_Y^2$.

%
%\begin{proposition}
%	\label{prop1}
%	Let $d_Y(t, t') = ||\Y_t-\Y_{t'}||$ and 
%	$r_Y(t, t_i) = d_Y(t, t_i) / d_Y(t, t_1)$, and $\{\widetilde Y_{t_i} : i=1, \ldots, d+1\}$ are the $d+1$ nearest neighbors to $\Y_t$. 
%	The CCM score of $Y_t$ on $X_t$ is 
%	\begin{align*}
%		\mathrm{CCM}(X_t \mid Y_t)
%		&=  \Big|\Big(X_0-\frac{\mu}{1-\alpha}\Big)\Big(\alpha^{t}-\sum_{i=1}^{d+1} w_Y(t_i, t)\alpha^{t_i}\Big) \\
%		&+ \Big(E_t -\sum_{i=1}^{d+1} w_Y(t_i, t)E_{t_i}\Big) \Big|,
%	\end{align*}
%	where we defined $E_{t}=\sum_{s=1}^{t}\alpha^{t-s}\epsilon_s$, and 
%	$w_Y(t_i,t) = e^{-r_Y(t, t_i)} / \sum_{j=1}^{d+1}e^{-r_Y(t, t_j)}.$
%	%%
%	Similarly, 
%	\begin{align*}
%	&\mathrm{CCM}(Y_t \mid X_t)  \\
%	&= \Big|\beta \Big(X_0-\frac{\mu}{1-\alpha}\Big)\Big(\alpha^{t-1}- \sum_{i=1}^{d+1} w_X(t_i', t)\alpha^{t_i'-1}\Big) + \\
%		&\beta\Big(E_t -\sum_{i=1}^{d+1} w_X(t_i', t) E_{t_i'}\Big) 
%		+ \Big(\zeta_t -\sum_{i=1}^{d+1} w_X(t_i', t)\zeta_{t_i'}\Big) \Big|,		
%	\end{align*} 
%	where $w_X(t_i',t) = e^{-r_X(t, t_i')} / \sum_{j=1}^{d+1}e^{-r_X(t, t_j')}.$
%\end{proposition}

\begin{assumption}\label{asp1}
Fix $t$ and let $L\to\infty$. Suppose that:
		\vspace{-10px}
	\begin{enumerate}
		\item[(a)] $\min_{i=1, \ldots, d+1} \min\{t_i', t_i \} \to \infty$;
		\vspace{-5px}
		\item[(b)] for both crossmaps, $\max_i | w_Y(t_i, t) - \frac{1}{d+1} |\to 0$, 
and $\max_i | w_X(t_i', t) - \frac{1}{d+1} | \to 0$;
		\vspace{-5px}
		\item[(c)] $\min_{i\neq j}{|t_i-t_j|} \to\infty$ and  $\min_{i\neq j}{|t_i'-t_j'|} \to\infty$.
	\end{enumerate}
\end{assumption}
\vspace{-5px}
{\bf Remarks.} Assumption 1(a) requires some form of smoothness for the delay-coordinate system, and is mild. Assumption 1(b) is similar to stationarity as it implies 
exchangeability within the sets $\{t_i: i=1, \ldots, d+1\}$ and $\{t_i': i=1, \ldots, d+1\}$.
Assumption 1(c) may be strict. It could fail, for instance, when the order statistics~(e.g., $\{\widetilde Y_{t_i}\}$) are periodic. 
% In the Supplement, we show that the assumptions apparently hold in our context.

\begin{theorem}\label{thm1}
Suppose that Assumptions~\ref{asp1}(a)-(c) hold for the CCM scores of 
the autoregressive model in Equation~\eqref{eq:AR}, then
\begin{align*}
\mathrm{CCM}(X_t \mid Y_t) & \xrightarrow{\text{d}} \mathrm{FN}\bigg( (X_0-\frac{\mu}{1-\alpha})\alpha^{t}, \frac{2-\alpha^{2t}}{1-\alpha^{2}}\sigma_X^2 \bigg), \\
\mathrm{CCM}(Y_t \mid X_t) &\xrightarrow{\text{d}} \mathrm{FN}\bigg( \beta(X_0-\frac{\mu}{1-\alpha})\alpha^{t-1}, \\ &\frac{2-\alpha^{2t}}{1-\alpha^{2}}\beta^2\sigma_X^2 + 2\sigma_Y^2 \bigg),
\end{align*}
where $\mathrm{FN}(\mu, \sigma^2) = |N(\mu, \sigma^2)|$ is the folded normal distribution with mean $\mu$ and variance $\sigma^2$.
\end{theorem}

\noindent\textbf{Remarks.} To unpack this theoretical result, we make the following remarks:
\begin{compactenum}[(a)]	
	\item When $\beta=0$, the dependence of $Y_t$ on $X_t$ is entirely lost, and so $X_t$ and $Y_t$ evolve independently implying that there is no driving factor in the system.
	CCM captures this relationship, since $\ccmXY = O_P(\sigma_X)$ and 
	$\ccmYX = O_P(\sigma_Y)$, with the two scores being independent~(this is shown in the proof of the theorem in the Supplement).
	
	\vspace{1mm}		
	\item When $\beta$ is small or moderately large, $Y_t$ is weakly dependent on $X_t$. 
	On average, we expect to see that $\ccmYX > \ccmXY$. CCM analysis indicates correctly that the driving factor in the system is $X_t$ and not $Y_t$~(recall that 
	we are using the absolute error-CCM, and so smaller values are better). 
	
	\vspace{1mm}	
	\item When $\beta$ is very large we could sometimes have $\ccmYX < \ccmXY$, which leads to the wrong ``causal direction''. This shows some inherent limitations of CCM, as 
	it depends to some extent on predictive ability, and so it can fail in similar 
	ways as Granger causality.
	% This is called ``synchrony" phenomenon~\citep{Sugihara2012}, which means that with extremely strong forcing, the reversed direction convergence can occur. ~\cite{Sugihara2012} suggests precluding ``synchrony" in CCM analysis.
%	\item When $\sigma_X=0$, $X_t$ is essentially deterministic, and converges fast to $\mu/(1-\alpha)$.\todo{how is this shown in Thm 1?} In contrast, $Y_t$ behaves increasingly as a random walk. Thus, $Y_t$ contains no information about $X_t$. The CCM analysis confirms this result, since it implies, in the limit, that $\ccmXY\approx 0$\todo{why?}, whereas $\ccmYX = O_P(\sigma_Y)$.
\end{compactenum}
In conclusion, Theorem~\ref{thm1} is a new connection of statistics and nonlinear dynamics. As mentioned earlier, the goal is not to analyze AR(1) per se, which indeed is a simple model, but to understand CCM's causal predictions by comparing to AR(1) coefficients. The theorem explains how and why CCM is capturing the directions correctly, and justifies using CCM in synthetic controls. A similar analysis of CCM in more complex models would be desirable, but is generally hard since CCM is highly nonlinear. We leave this for future work.

\subsection{CCM+SCM Method and Proposition 99}
\label{sec:procedure}
\begin{figure}[!t]
	\centering
	\includegraphics[width=0.45\textwidth]{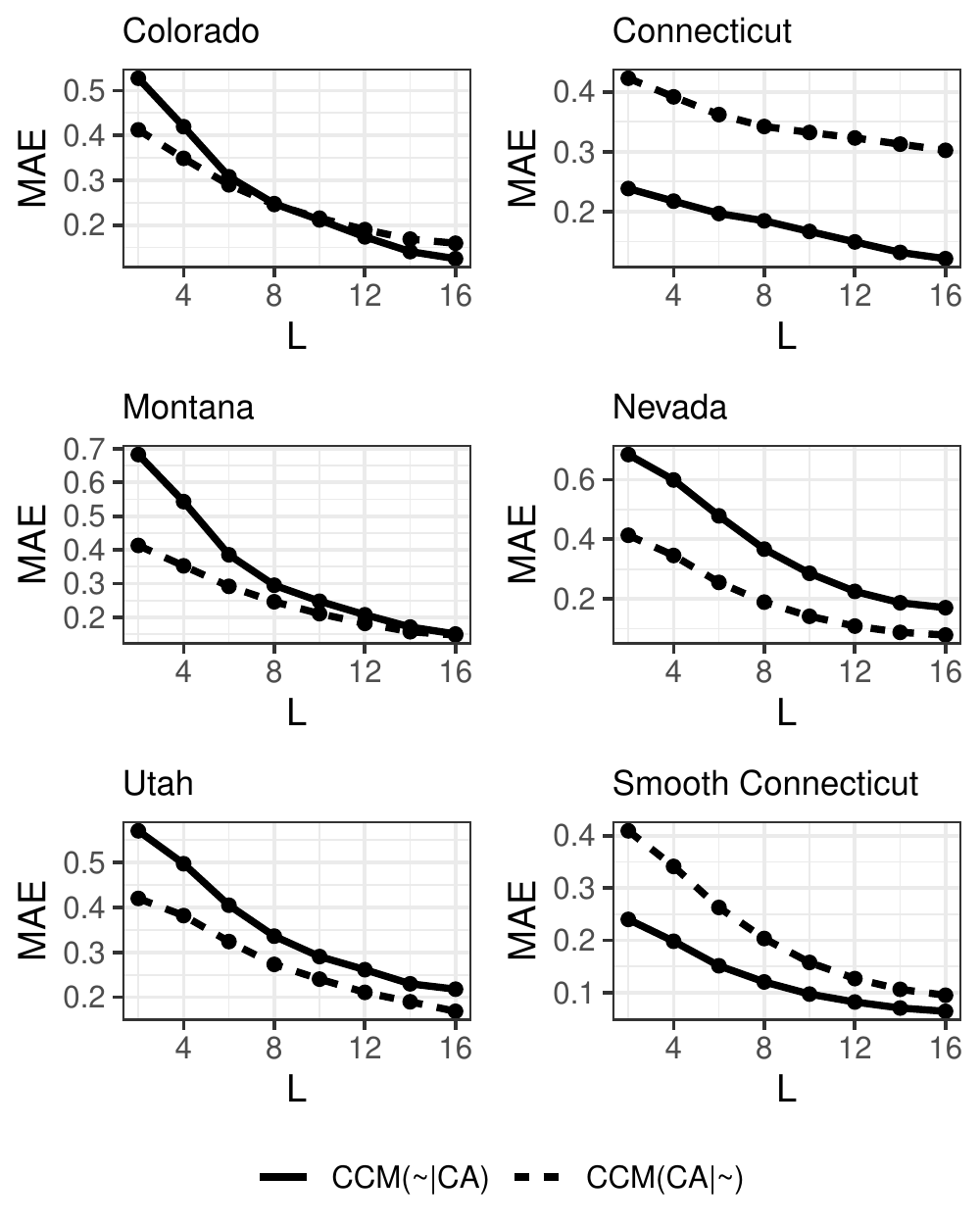}
	\vspace{-0.1in}\caption{CCM scores based on MAE between California and control states with varying library size $L$. $\mathrm{CCM}(CA \mid\sim)$ means cross predicting California from control, and $\mathrm{CCM}(\sim\mid CA)$ means cross predicting control from California. }
	\label{fig4}  
	\vspace{-10pt}
\end{figure}

Here, we present CCM scores in the Proposition 99 example introduced in Section~\ref{sec:motivation}. 
Specifically, California (CA) is cross-mapped with five control states selected by the standard synthetic control method as shown in Equation~\eqref{eq:ca_example1}. We use per-capita cigarette sales  from the pre-intervention period as the outcome variable, which gives 19 data points for each 
unit's time-series. The cross predictability measured by the CCM scores for each pair is shown in Figure~\ref{fig4}. For example, the California-Colorado pair includes two CCM curves, namely,
$\mathrm{CCM}(Y_{\mathrm{CA}, t}\mid Y_{\mathrm{CO}, t})$ and $\mathrm{CCM}(Y_{\mathrm{CO}, t}\mid Y_{\mathrm{CA}, t})$.

We see that cross predictability for all pairs roughly converge as the library size, $L$, grows. Furthermore, most pairs converge to the same low level of CCM score, indicating a strong and bidirectional dynamical relationship between the state pairs. The only obvious exception is the California-Connecticut pair, where a big gap occurs between the two curves exists, indicating a weak dynamical relationship between them. 

In particular, we see that Connecticut is better predicted from California than the other way round. For this reason, we argue that Connecticut is not a suitable control for California and should be removed from the donor pool. If we apply an averaging transformation to smooth out the 1970-1980 trend of Connecticut, the CCM score changes and now shows a strong dynamic coupling between the two states (bottom-right plot in Figure~\ref{fig4}). If Connecticut is removed, SCM will pick Minnesota. However, CCM will screen Minnesota as well because the cigarette price trends are similar between Minnesota and Connecticut, but distinct from California. 

\begin{figure*}[t!]
	\centering
	\begin{minipage}[b]{0.4\linewidth}
		\includegraphics[width=1\textwidth]{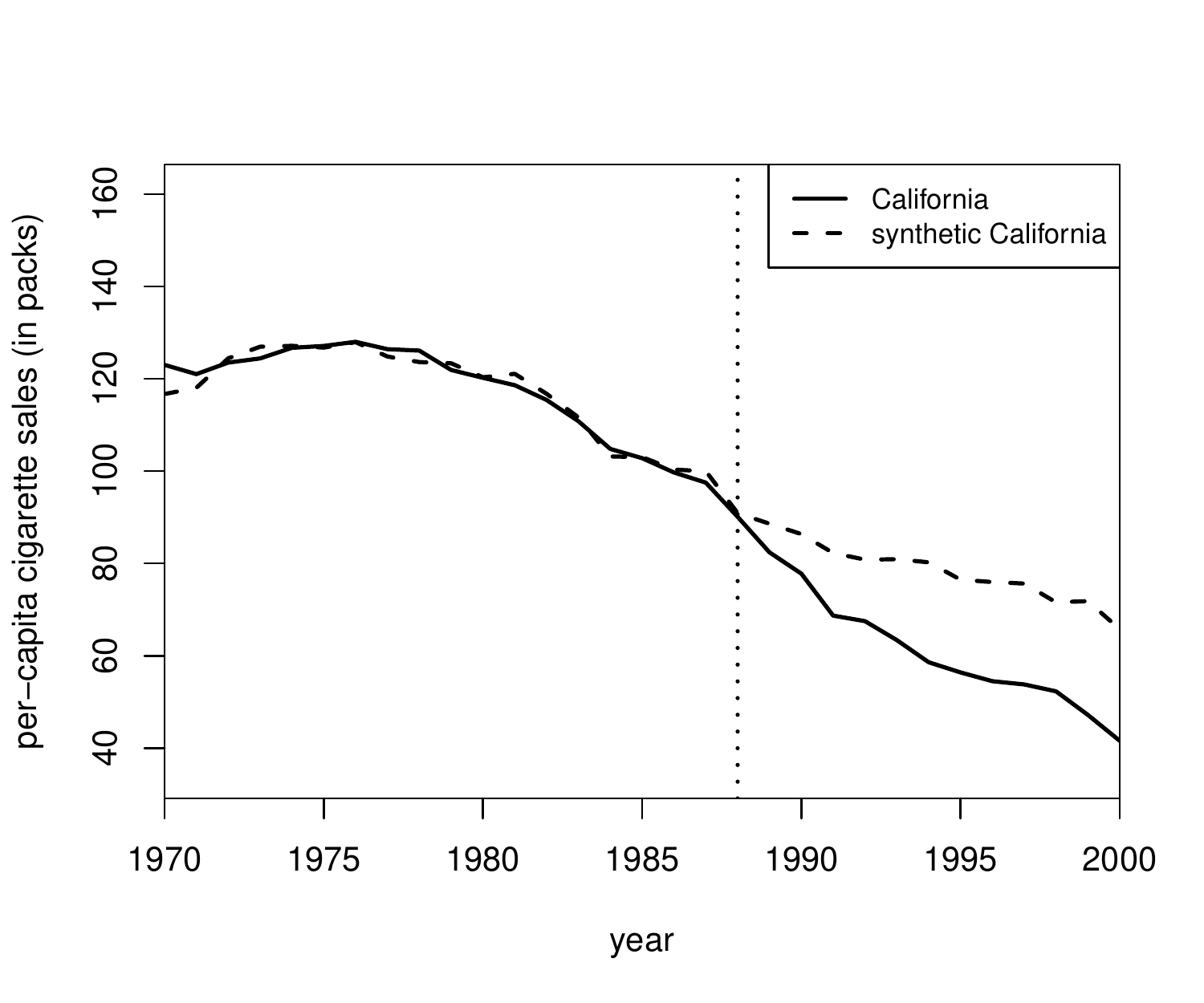}  
	\end{minipage}
	\begin{minipage}[b]{0.4\linewidth}
		\includegraphics[width=1\textwidth]{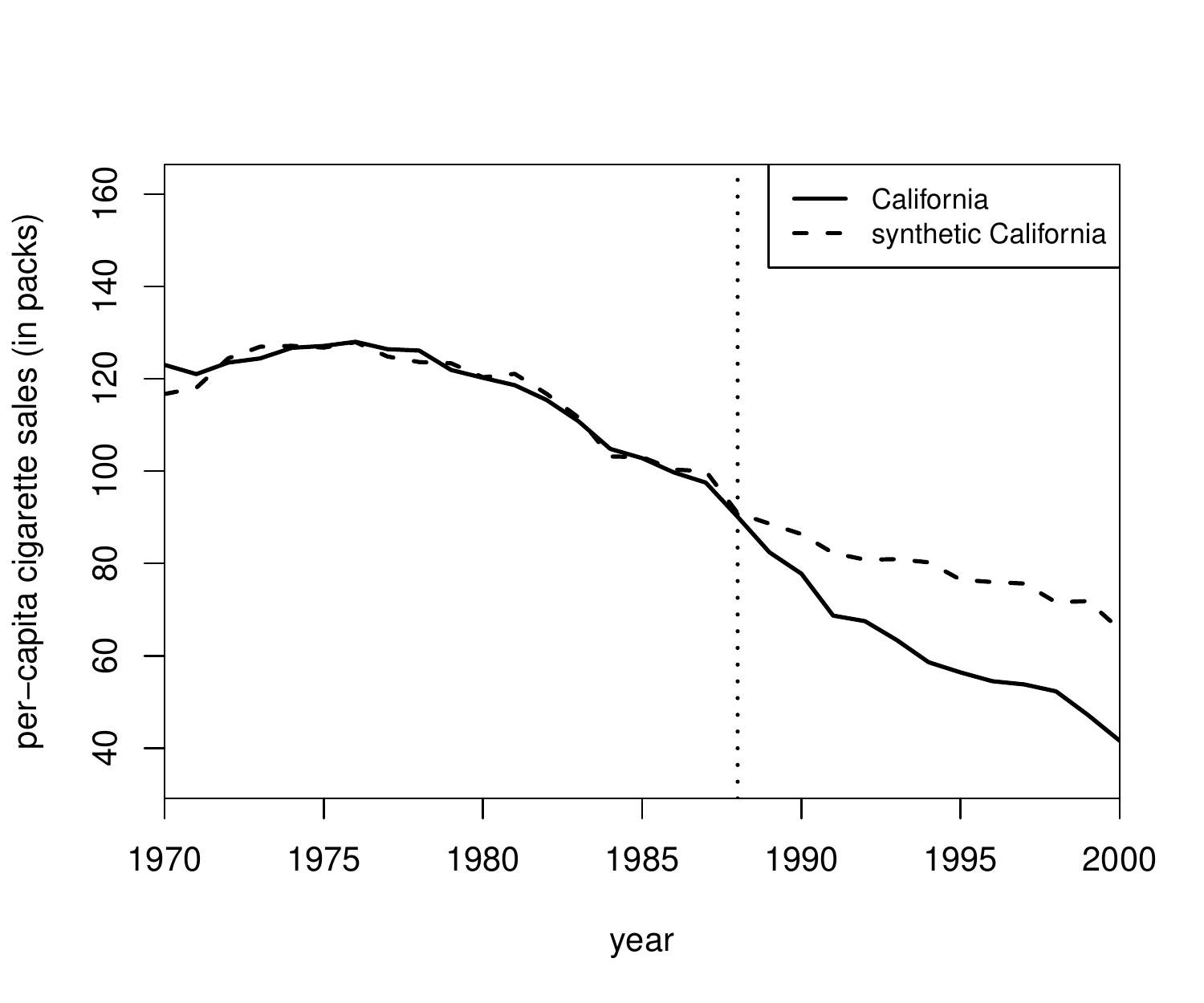}
	\end{minipage}
	\vspace{-10pt}
	\caption{Trends in per-capita cigarette sales \textbf{after CCM pre-screening}. The solid line is actual California and the dashed line is synthetic California, while the vertical line indicates time of intervention. \textbf{Left}: original setting. \textbf{Right}: adversarial setting where artificial units are added to the donor pool.}\label{fig2}  
\end{figure*}

Our proposed method is therefore to use CCM to filter out controls that have a weak 
dynamical relationship with the treated unit, and then apply the standard SCM method as described in Section~\ref{sec:prelims}. We refer to this method as ``\CCMSCM". In practice, we propose that \CCMSCM\ filters out a control unit if in the two CCM plots with the treated unit, either the minimum MAE or the MAE gap exceed some thresholds.
%
% The optimal embedding dimension is selected by cross validation based on the correlation between observed and predicted values. The library size only depends on the length of time series data, and up to the point when CCM converges. Thus, it does not depend on the number of control units. 
To determine the cutoff values, we may use Monte-Carlo simulations 
where we add noise to the original series, and then estimate the null distribution of CCM values 
under a hypothesis of weak dynamical relationship. 
% We run the simulation between Connecticut and California as shown in Figure~\ref{fig:gap}, and determine filtering condition that either the minimum MAE is greater than 0.1 in both direction, or the MAE gap is greater than 0.1. As such, Connecticut is filtered out due to the weak dynamical relationship with California, and it is evident that the plots for Connecticut-California are distinct from all others in Figure~\ref{fig4}.

%For example, in Figure~\ref{fig4} the CCM plots for Connecticut-California have a gap that is greater than 0.1 at $L=16$, and so we filter out Connecticut. These cutoffs are necessarily ad hoc at the moment, since there is currently no theory for a more data-adaptive selection. We chose the aforementioned numerical values based on our extensive applied studies with CCM. Even though a more principled cutoff selection would clearly be desirable, choosing appropriate cutoffs has not been a critical issue in the applications we considered. For instance, in Figure~\ref{fig4} it is evident that the plots for Connecticut-California are distinct from all others, and thus Connecticut stands out as a state with weak dynamical relationship with California.

%\begin{figure}[t!]
%	\centering
%	\includegraphics[width=0.49\textwidth]{aistats_gap.pdf}
%	\vspace{-0.3in}\caption{CCM score and gap distributions between Connecticut and California.}
%	\label{fig:gap}  
%	\vspace{-10pt}
%\end{figure}

To illustrate the potential of \CCMSCM, we return to the 
example of Section~\ref{sec:motivation}, where we showed that adversarial units in the control pool affected the performance of synthetic controls.
Figure~\ref{fig2} shows that \CCMSCM\ is able to screen 
the adversarial units, and is able to produce a synthetic control that is indistinguishable from the non-adversarial setting. In the following section, we explore the performance of \CCMSCM\ further through simulated studies and real-world data.

\section{Experiments and Applications}
\label{sec:experiment}
%  (10/7) was badly written
% Since we just showed that \CCMSCM\ performs well without changing treatment effect in the normal setting, we investigate further how CCM behaves in the worse settings. To do so, we create the adversarial settings where artificial units are added to the donor pool to bias the synthetic control selection procedure. Our goal is to show that CCM is a general framework for pre-screening that applies to different settings and implementations.

Here, we design adversarial settings where artificial units are added to the donor pool to bias the synthetic control method. Of particular interest is whether \CCMSCM\ can help filter out the artificial units in all cases, 
without affecting the baseline performance when artificial units are not present. We also consider real-world applications.

\subsection{Simulations with Artificial Units}
First, we expand the tobacco legislation example of Section~\ref{sec:motivation}  by introducing a larger set of artificial units, which are created adversarially. These artificial data were created based on real-world time series macroeconomic data. The detailed generation process can be found in the Supplement.
%As a sanity check, Figure~\ref{fig:syndata} shows that despite the variation these artificial units look generally similar to the true units. 
We run simulation studies in which the true effect is known for the treated unit. We replace the true California with the following formula:
\begin{align*}
\widetilde{Y}_{\mathrm{CA}, t} &= 0.164 \times Y_{\mathrm{CO}, t} + 0.069 \times Y_{\mathrm{CT}, t} + 0.199 \times Y_{\mathrm{MO}, t} \\
&+ 0.234\times Y_{\mathrm{NV}, t} +
0.334 \times Y_{\mathrm{UT}, t} +\tau \mathbb{I}\{t \geq 1989\},
\end{align*}
where $\tau$ is the true treatment effect and the other terms construct the synthetic California from the original data. This construction ensures that the ground truth of synthetic California in the post-intervention period is known. To illustrate that CCM is a general framework for pre-screening, we incorporate CCM to three different synthetic control methods: 1) \texttt{SCM}: vanilla synthetic control method by~\cite{Abadie2010}; 2) \texttt{MC}: matrix completion method for causal panel data by~\cite{athey2017matrix}; 3) \texttt{RSC}: robust synthetic control method by~\citet{amjad2018robust}. 
%

%\begin{figure}[t!]
%	\centering
%	\includegraphics[width=0.49\textwidth]{aistats_syndata.pdf}
%	\vspace{-0.3in}\caption{Artificial and true data in Tobacco case.}
%	\label{fig:syndata}  
%	\vspace{-10pt}
%\end{figure}

\begin{figure*}[h]
	\centering
	\includegraphics[width=0.95\textwidth]{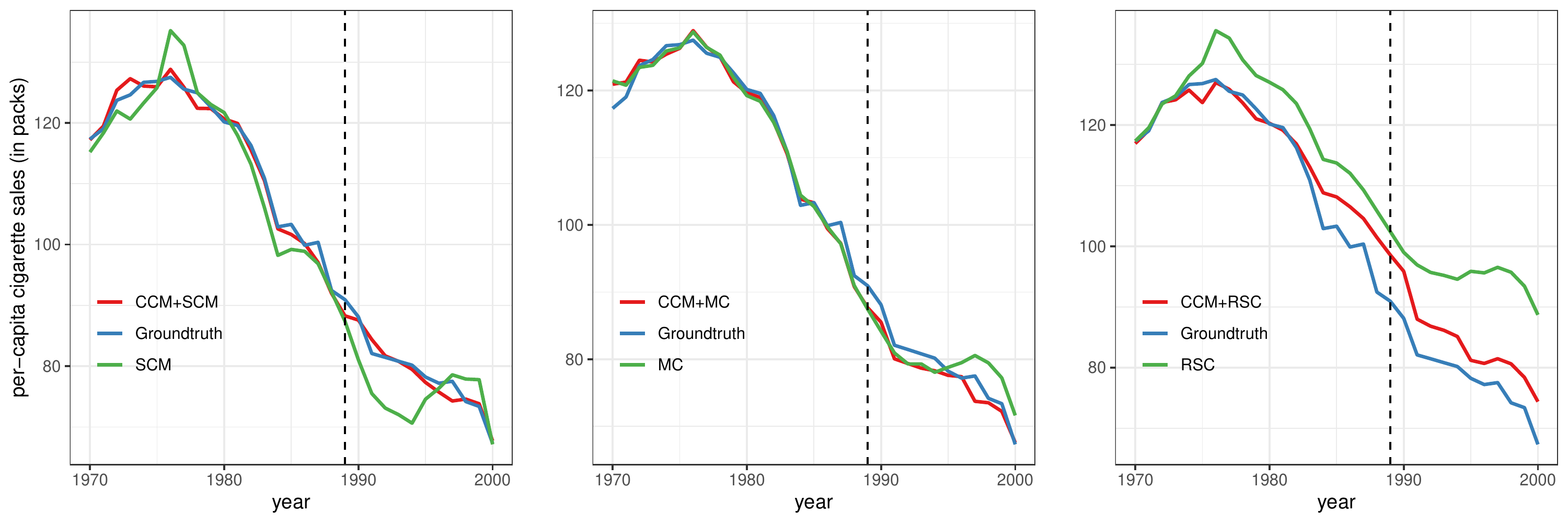}
	\vspace{-0.1in}\caption{Comparisons of between ground truth and constructed synthetic California with and without CCM by three synthetic control methods.
	(In this simulation, $\tau=4$, but the results are robust to this value).}
	\label{fig5}  
	\vspace{-0.1in}
\end{figure*}

The comparisons between the ground truth and three synthetic control methods with and without CCM are visualized in Figure~\ref{fig5}. The different methods lead to the same conclusion: with CCM pre-screening the outcome estimates, in both pre- and post-intervention periods, are closer to the ground truth than the original method alone. 
We note that \texttt{MC} works by matrix completion instead of selecting control units so \texttt{CCM+MC} behaves very similarly to \texttt{MC}. \texttt{RSC} alone does not work well because it uses the artificial unit to construct the synthetic control, and the denoising via singular value thresholding does not help here. The result suggests that CCM can help in synthetic control models, and is robust to the selection of the underlying outcome imputation model. Intuitively, this is  because CCM is able to capture nonlinear dynamical information that is not captured by standard statistical models.

\subsection{Real-World Applications}
Here, we consider how many artificial units CCM is able to filter out with real-world data. To that end, we work with two real applications: one is still the California's Tobacco Control Program studied in~\cite{Abadie2010}; and the other is on the economic costs of the Brexit referendum vote on UK's GDP reported in~\cite{Born2017}. We only compare SCM and \CCMSCM. Our results are robust to the selection of the underlying model. As before, artificial units are created from noisy copies of real-world time series. 

\begin{figure}[]
	\centering
	\includegraphics[width=0.48\textwidth]{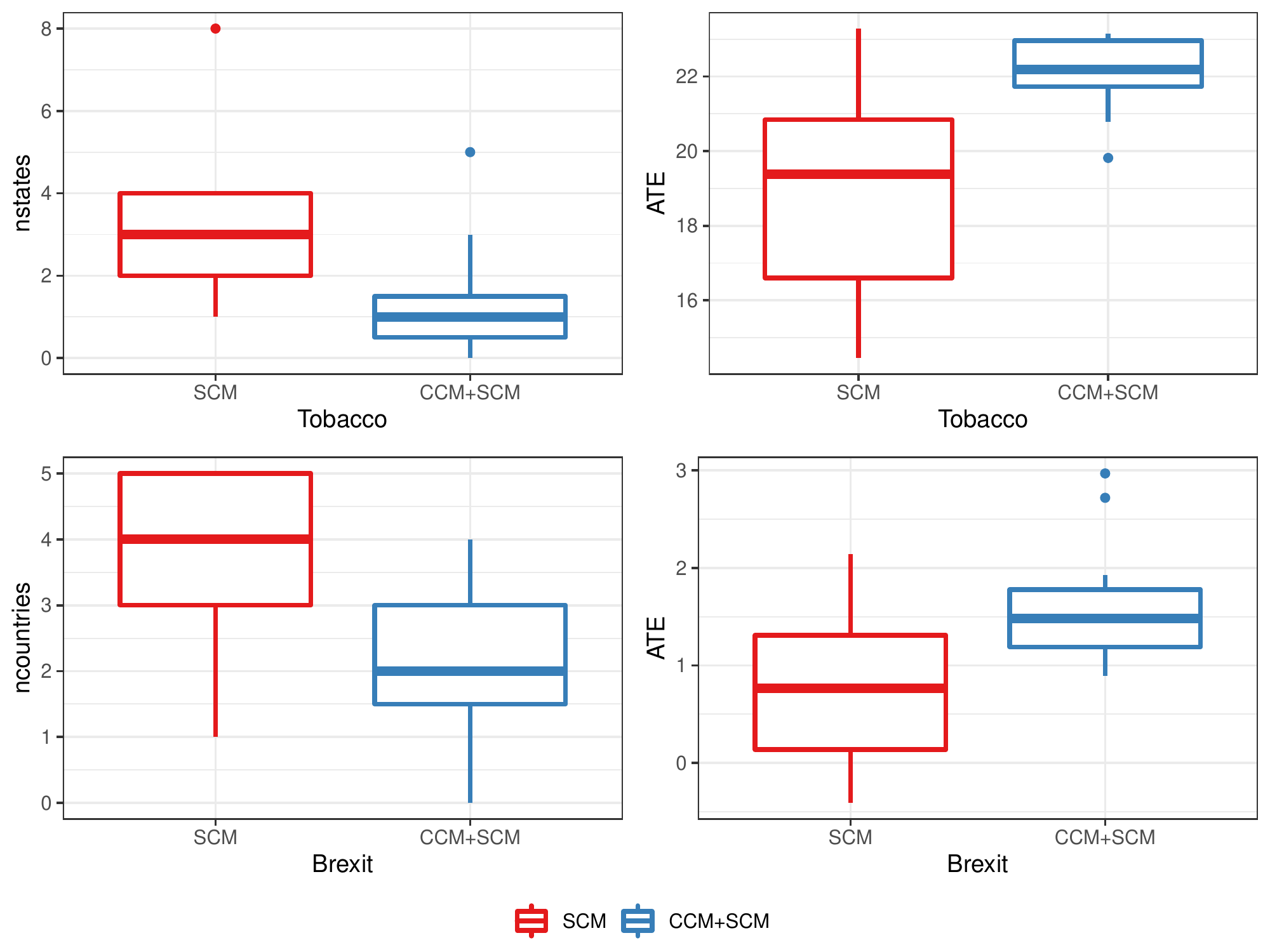}
	\vspace{-0.3in}\caption{Box plots of the number of artificial control units selected by each method~(left), namely SCM and \CCMSCM, and average treatment effect~(right) in the \texttt{Tobacco} and \texttt{Brexit} applications.}
	\label{fig:art}  
%	\vspace{-10pt}
\end{figure}
Figure~\ref{fig:art} shows the number of artificial units that are selected as controls, and the corresponding average treatment effect (ATE). Clearly,~\CCMSCM\ selects much fewer artificial units than just SCM. In addition, \CCMSCM\ generates more stable estimates of ATE. Specifically, in the tobacco example, the ATE reported by \CCMSCM\ remains stable around 20, which is close to the original ATE ($\approx 19$~packets) estimated by SCM without artificial controls. In contrast, the ATE estimate from SCM with artificial controls is more varied in a range from 5 to 23. Another interesting phenomenon is that the ATE estimate from SCM can be {\em negative} under certain artificial controls in \texttt{Brexit}. This means that the effects of the Brexit vote may have been overstated in ongoing econometric work that uses synthetic control methods, as the estimates are likely to be sensitive to control pool construction.

\section{Discussion}
\label{sec:conclude}
%  We proposed to screen out control units that are weakly related to treated units, and illustrated CCM through a simple autoregression model. Empirically, we showed on both simulated data and real applications that CCM filter can mitigate bias from ``cherry-picking'' of control units, and does not depend on the particular SCM model used. 
In this section, we discuss some general aspects of our work, particularly CCM as a causal inference method, and its underlying assumptions.
 
Our first point revolves around the use of CCM, and related dynamical systems methods, 
for causal inference.  In particular, due to the success of CCM in quantifying dynamical relationships illustrated here, it may be tempting to consider CCM as a general method for causal inference~\citep{Sugihara2012,Deyle6430}. However, we do not advocate doing that directly, for two main reasons. First, the statistical properties of methods such as CCM are not well known. Theorem~\ref{thm1} in this paper is a step to this direction, but more work is needed. 
Second, CCM does not account for the observation model (e.g., the treatment assignment mechanism), which is crucial in causal inference~\citep{imbens2015causal}.
We also provide counterexamples and additional discussion  in the Supplement.

% We also need to clarify that that even through CCM+SCM was developed here as a mechanistic way to filter out control units, it is not meant  a replacement of expert knowledge. envision our method as a safeguard against cherrypicking or any spurious extrapolations from dynamically unrelated control units, and definitely not as a replacement of expert knowledge.

Regarding the stationarity assumption in Theorem~\ref{thm1}~(i.e., $|\alpha|<1$, and Assumption~1(b)), 
we note that CCM assumes a deterministic (possibly chaotic) system, and so stationarity is roughly mapped to ``evolution on an attractor''. Stationarity in Theorem~\ref{thm1} is thus not directly applicable to Takens' theorem, and is assumed only in our effort to understand CCM vis-à-vis AR(1). In this work, we argue that CCM can strengthen the model continuity assumption (pre- and post-treatment), which is left implicit in synthetic controls. But there is much left to understand about the connection between chaotic and stochastic systems~\citep{casdagli1992chaos}. We hope that this work provides a good motivation.

% Furthermore, CCM provides a viable way to connect structural and reduced-form approaches. It would be interesting to know, for example, under what conditions reduced-form methods could use CCM to control for hidden economic structure.

\section{Conclusion}
In this paper, we leveraged results from dynamical systems theory to quantify the strength of dynamic relationship between treated and control units in causal inference. We showed that this is useful in the context of comparative cases studies to guard against cherry-picking of potential controls, which is an important concern in practice.

More generally, our work opens up the potential for an interplay between dynamical systems theory and causal inference. In practice, interventions typically occur on complex dynamical systems, such as an auction or a labor market, which always evolve, before and after treatment.  Future work could focus more on theoretical connections between embedding methods, such as CCM, and standard treatment effects in econometrics, especially if we view the filtering process described in Section~\ref{sec:method} as a way to do treated-control matching. 

\section*{Acknowledgment}
We thank our anonymous reviewers for the feedback that improved this final version of the paper. We also thank Hao Ye, who provided assistance in understanding CCM. Y.D.'s work has been partially supported by the NSF (CCF-1439156, CCF-1823032, CNS-1764039).

\bibliography{ccm_syn}
\bibliographystyle{apalike}

\end{document}

\maketitle
\section{Proof Theorem 1}\label{sec:theory}

\begin{proposition}\label{prop1}
	Let $d_Y(t, t') = ||\Y_t-\Y_{t'}||$ and 
	$r_Y(t, t_i) = d_Y(t, t_i) / d_Y(t, t_1)$, where $\Y_t=[Y_t, Y_{t-\tau},\ldots, Y_{t-(d-1)\tau}]$ is the delayed coordinate embedding of $Y_t$, and $\{\widetilde Y_{t_i} : i=1, \ldots, d+1\}$ are the $d+1$ nearest neighbors to $\Y_t$. 
	The CCM score of $Y_t$ on $X_t$ based on MAE metric is equal to:
	\begin{align*}
	\mathrm{CCM}(X_t \mid Y_t)
	&=  \Big|\Big(X_0-\frac{\mu}{1-\alpha}\Big)\Big(\alpha^{t}-\sum_{i=1}^{d+1} w_Y(t_i, t)\alpha^{t_i}\Big) + \Big(E_t -\sum_{i=1}^{d+1} w_Y(t_i, t)E_{t_i}\Big) \Big|,
	\end{align*}
	where $E_{t}=\sum_{s=1}^{t}\alpha^{t-s}\epsilon_s$, and 
	$
	w_Y(t_i,t) = e^{-r_Y(t, t_i)} / \sum_{j=1}^{d+1}
	e^{-r_Y(t, t_j)}.$
	Similarly, let $d_X(t, t') = ||\X_t-\X_{t'}||$ and 
	$r_X(t, t_i') = d_X(t, t_i') / d_X(t, t_1')$, where $\X_t=[X_t, X_{t-\tau},\ldots, X_{t-(d-1)\tau}]$ is the delayed coordinate embedding of $X_t$, and $\{\widetilde X_{t_i'} : i=1, \ldots, d+1\}$ are the $d+1$ nearest neighbors to $\X_t$. 
	The CCM score of $X_t$ on $Y_t$ based on MAE metric is equal to:
	\begin{align*}
	\mathrm{CCM}(Y_t \mid X_t)  
	= \Big|\beta \Big(X_0-\frac{\mu}{1-\alpha}\Big)\Big(\alpha^{t-1}- \sum_{i=1}^{d+1} w_X(t_i', t)\alpha^{t_i'-1}\Big) + 
	\beta\Big(E_t -\sum_{i=1}^{d+1} w_X(t_i', t) E_{t_i'}\Big) 
	+ \Big(\zeta_t -\sum_{i=1}^{d+1} w_X(t_i', t)\zeta_{t_i'}\Big) \Big|,	
	\end{align*} 
	where $w_X(t_i',t) = e^{-r_X(t, t_i')} / \sum_{j=1}^{d+1}
	e^{-r_X(t, t_j')}.$
\end{proposition}

\begin{proof}
	Recall that the AR model is as follows:
	\begin{align*}
	X_t & = \alpha X_{t-1} + \mu + \epsilon_t, \\
	Y_t & = \beta X_{t-1} + \mu + \zeta_t,
	\end{align*}
	where $\alpha$, $\beta$ are fixed, with $|\alpha|<1$, $\beta\geq 0$, $\mu$ is the drift, $\epsilon_t \sim \mathcal{N}(0, \sigma_X^2)$ and $\zeta_t \sim \mathcal{N}(0, \sigma_Y^2)$ are zero-mean and constant-variance normal errors. 
	We can solve the recursion to obtain:
	\begin{align*}
	X_t & = \alpha^t X_0 + P_t(\alpha) \mu + E_t,\\
	Y_t & = \beta \alpha^{t-1} X_0 + \left(\beta P_{t-1}(\alpha) + 1\right) \mu + \beta E_{t}  + \zeta_t,
	\end{align*}
	where $P_t(\alpha) = 1 + \alpha + \ldots + \alpha^{t-1} = (1-\alpha^t) / (1 - \alpha)$ and $E_{t}$ is the summation of weighted error terms $E_{t}=\sum_{s=1}^{t}\alpha^{t-s}\epsilon_s$.
%	It is obvious that $P_t(\alpha) \to 1/ (1-\alpha)$ as $t\to\infty$.
	
%Since $X$ drives $Y$ in this model, we can generate a cross estimation of $X_t$ from $Y_t$ denoted as $\hat{X}_t$. For every $\{t|t>L\}$, where $L$ is the library size, we can find $N_Y(t) = \{t_1, t_2, \ldots, t_{d+1}\}$ such that $\{ \Y_{t_i}|t_i\in N_Y(t),0<t_i\leq L\}$ are nearest neighbors of $\Y_t$, where $\Y_t$ is the corresponding  delayed coordinate embedding with dimension $d$. Thus, 
The cross estimation for $X_t$ is
	\begin{align}
	\hat X_t &=\sum_{i=1}^{d+1} w_Y(t_i, t) X_{t_i} 
	= \sum_{i=1}^{d+1} w_Y(t_i, t) \left(\alpha^{t_i} X_0 + P_{t_i}(\alpha) \mu + E_{t_i} \right),
	\nonumber
	\end{align}
		where $w_Y(t_i,t) = e^{-r_Y(t, t_i)} / \sum_{j=1}^{d+1}e^{-r_Y(t, t_j)}$. Then, the CCM score of $Y_t$ on $X_t$ based on MAE can be expressed as:
\begin{align} \label{eq:ccm_x}
	|X_t - \hat X_t| &= \Big|\alpha^t X_0 + P_t(\alpha) \mu + E_t - \Big(\sum_{i=1}^{d+1} w_Y(t_i, t)(\alpha^{t_i} X_0 + P_{t_i}(\alpha) \mu + E_{t_i})\Big)\Big| \nonumber \\
	&= \Big|X_0\Big(\alpha^{t}-\sum_{i=1}^{d+1} w_Y(t_i, t)\alpha^{t_i}\Big) + \mu\Big(P_{t}(\alpha)- \sum_{i=1}^{d+1} w_Y(t_i, t)P_{t_i}(\alpha) \Big) + \Big(E_t -\sum_{i=1}^{d+1} w_Y(t_i, t)E_{t_i}\Big) \Big| \nonumber \\
	&= \Big|X_0\Big(\alpha^{t}-\sum_{i=1}^{d+1} w_Y(t_i, t)\alpha^{t_i}\Big) + \mu\Big(\frac{1-\alpha^{t}}{1-\alpha}-\sum_{i=1}^{d+1} w_Y(t_i, t)\frac{1-\alpha^{t_i}}{1-\alpha} \Big) + \Big(E_t -\sum_{i=1}^{d+1} w_Y(t_i, t)E_{t_i}\Big) \Big| \nonumber \\
	&=\Big|\Big(X_0-\frac{\mu}{1-\alpha}\Big)\Big(\alpha^{t}-\sum_{i=1}^{d+1} w_Y(t_i, t)\alpha^{t_i}\Big) + \Big(E_t -\sum_{i=1}^{d+1} w_Y(t_i, t)E_{t_i}\Big) \Big|.
	\end{align}	
	
	Similarly, the cross estimation for $Y_t$ is
	\begin{align}
	\hat Y_t &=\sum_{i=1}^{d+1} w_X(t'_i, t) Y_{t'_i} 
	= \sum_{i=1}^{d+1} w_X(t'_i, t)(\beta \alpha^{t'_i-1} X_0 +(\beta P_{t'_i-1}(\alpha) + 1) \mu + \beta E_{t'_i} + \zeta_{t'_i} ),
	\end{align}
		where $w_X(t_i',t) = e^{-r_X(t, t_i')} / \sum_{j=1}^{d+1}e^{-r_X(t, t_j')}$. Then, the CCM score of $X_t$ on $Y_t$ based on MAE can be expressed as:
	\begin{align} \label{eq:ccm_y}
	|Y_t - \hat Y_t | 
	&= \Big|\beta \alpha^{t-1} X_0 + \Big(\beta P_{t-1}(\alpha) + 1\Big) \mu + \beta E_t +\zeta_t - \Big(\sum_{i=1}^{d+1} w_X(t'_i, t)(\beta \alpha^{t'_i-1} X_0 + (\beta P_{t'_i-1}(\alpha) + 1) \mu + \beta E_{t'_i} + \zeta_{t'_i} )\Big)\Big|  \nonumber \\
	&= \Big|\beta X_0\Big(\alpha^{t-1}-\sum_{i=1}^{d+1} w_X(t'_i, t)\alpha^{t'_i-1}\Big) + \beta\mu\Big(P_{t-1}(\alpha)- \sum_{i=1}^{d+1} w_X(t'_i, t)P_{t'_i-1}(\alpha) \Big) +\beta\Big(E_t -\sum_{i=1}^{d+1} w_X(t'_i, t)E_{t'_i}\Big) \nonumber \\ 
	& + \Big(\zeta_t -\sum_{i=1}^{d+1} w_X(t'_i, t)\zeta_{t'_i}\Big) \Big|  \nonumber \\
	&= \Big|\beta X_0\Big(\alpha^{t-1}-\sum_{i=1}^{d+1} w_X(t'_i, t)\alpha^{t'_i-1}\Big) + \beta\mu\Big(\frac{1-\alpha^{t-1}}{1-\alpha}-\sum_{i=1}^{d+1} w_X(t'_i, t)\frac{1-\alpha^{t'_i-1}}{1-\alpha} \Big) +\beta\Big(E_t -\sum_{i=1}^{d+1} w_X(t'_i, t)E_{t'_i}\Big) \nonumber \\ 
	& + \Big(\zeta_t -\sum_{i=1}^{d+1} w_X(t'_i, t)\zeta_{t'_i}\Big) \Big|  \nonumber \\
	&= \Big|\beta \Big(X_0-\frac{\mu}{1-\alpha}\Big)\Big(\alpha^{t-1}-\sum_{i=1}^{d+1} w_X(t'_i, t)\alpha^{t'_i-1}\Big) +\beta\Big(E_t -\sum_{i=1}^{d+1} w_X(t'_i, t)E_{t'_i}\Big) + \Big(\zeta_t -\sum_{i=1}^{d+1} w_X(t'_i, t)\zeta_{t'_i}\Big) \Big|.
	\end{align} 
\end{proof}

We now proceed to the proof of Theorem 1 using the results in Equation~\eqref{eq:ccm_x}
and Equation~\eqref{eq:ccm_y}. 
We repeat the Assumption 1 in the main paper.

\begin{assumption}\label{asp1}
	For the CCM scores in Proposition~\ref{prop1}, fix $t$ and let $L\to\infty$,
	and suppose that:
	\begin{compactenum}[(a)]
		\item $\min_{i=1, \ldots, d+1} \min\{t_i', t_i \} \to \infty$;
		\item $\lim\sup_{i=1, \ldots, d+1} | w_Y(t_i, t) - \frac{1}{d+1} | = 0$, 
		and~$\lim\sup_{i=1, \ldots, d+1} | w_X(t_i', t) - \frac{1}{d+1} | = 0$;
		\item $\min_{i\neq j}{|t_i-t_j|} \to\infty$ and  $\min_{i\neq j}{|t_i'-t_j'|} \to\infty$.
	\end{compactenum}
\end{assumption}

\begin{theorem}\label{thm1}
	Suppose that Assumptions~\ref{asp1}(a)-(c) hold for the CCM scores in Proposition~\ref{prop1}, then
	\begin{align*}
	\mathrm{CCM}(X_t \mid Y_t) & \xrightarrow{\text{d}} \mathrm{FN}\bigg( (X_0-\frac{\mu}{1-\alpha})\alpha^{t}, \frac{2-\alpha^{2t}}{1-\alpha^{2}}\sigma_X^2 \bigg), \\
	\mathrm{CCM}(Y_t \mid X_t) &\xrightarrow{\text{d}} \mathrm{FN}\bigg( \beta(X_0-\frac{\mu}{1-\alpha})\alpha^{t-1},
	\frac{2-\alpha^{2t}}{1-\alpha^{2}}\beta^2\sigma_X^2 + 2\sigma_Y^2 \bigg),
	\end{align*}
	where $\mathrm{FN}(\mu, \sigma^2) = |N(\mu, \sigma^2)|$ is the folded normal distribution with mean $\mu$ and variance $\sigma^2$.
\end{theorem}

\begin{proof}
From $\alpha < 1$ and Assumption 1(a) we get:
\begin{align*}
\sum_{i=1}^{d+1} w_Y(t_i, t)\alpha^{t_i} \le \sum_{i=1}^{d+1} w_Y(t_i, t) \alpha^{\min_i\min\{t_i, t_i'\}}
\le   \alpha^{\min_i\min\{t_i, t_i'\}} \sum_{i=1}^{d+1} w_Y(t_i, t)
 =  \alpha^{\min_i\min\{t_i, t_i'\}} \to 0.
\end{align*}
Since $E_t = \sum_{s=1}^{t}\alpha^{t-s}\epsilon_s =\alpha^{t-1}\epsilon_1 + \alpha^{t-2}\epsilon_1+ \cdots + \epsilon_t$ and $\epsilon_t \sim \mathcal{N}(0, \sigma_X^2)$, we have 
$$
\Ex(E_t) = 0,~\text{and}~\var(E_t) = \frac{1-\alpha^{2t}}{1-\alpha^2}\sigma_X^2.
$$

For $t_i \neq t_j$, it is straightforward to show that
$
\cov(E_{t_i}, E_{t_j}) = O(\alpha^{|t_i-t_j|})  \to 0,
$
where the limit follows from Assumption 1(c). Similarly, the results hold for $t_i', t_j'$.
$$
\Ex(E_{t_i})~\text{and}~
\text{Cov}(E_{t_i},E_{t_i})=\frac{1-\alpha^{2t_i}}{1-\alpha^{2}}\sigma_X^2  \to \frac{1}{1-\alpha^{2}}\sigma_X^2 .
$$
From Assumption 1(b) we have:
\begin{align*}
\sum_{i=1}^{d+1} w_Y(t_i, t)E_{t_i} \toD \frac{1}{d+1} \sum_{i=1}^{d+1} E_{t_i}\
\to N(0, \frac{\sigma_X^2}{1-\alpha^2}),
\end{align*}
from which it follows that
$$
E_t -\sum_{i=1}^{d+1} w_Y(t_i, t)E_{t_i}\sim N(0, \frac{2-\alpha^{2t}}{1-\alpha^{2}}\sigma_X^2).
$$
Hence,  $\mathrm{CCM}(X_t \mid Y_t)$ converges in a distribution to
\begin{align*}
\mathrm{CCM}(X_t \mid Y_t) & \xrightarrow{\text{d}} \mathrm{FN}\bigg( (X_0-\frac{\mu}{1-\alpha})\alpha^{t}, 
\frac{2-\alpha^{2t}}{1-\alpha^{2}}\sigma_X^2 \bigg),
\end{align*}
where $\mathrm{FN}(\cdot,\cdot)$ is the folded normal distribution.

Similarly, since $\zeta_t \sim \mathcal{N}(0, \sigma_Y^2)$ and $\lim\sup_{i=1, \ldots, d+1} | w_X(t_i', t) - \frac{1}{d+1} | = 0$, we have
\begin{align*}
\sum_i w_X(t_i', t) \zeta_{t_i'} = \frac{1}{d+1}\sum_{i=1}^{d+1} w_X(t'_i, t)\zeta_{t'_i}
+ o_P(1) \toD N(0, \sigma_Y^2).
\end{align*}
It follows that
$\zeta_t -\sum_{i=1}^{d+1} w_X(t'_i, t)\zeta_{t'_i}\sim \mathcal{N}(0, 2\sigma_Y^2)$,
and that $\mathrm{CCM}(Y_t \mid X_t)$ converges in a distribution to
\begin{align*}
\mathrm{CCM}(Y_t \mid X_t) &\xrightarrow{\text{d}} \mathrm{FN}\bigg( \beta(X_0-\frac{\mu}{1-\alpha})\alpha^{t-1}, \frac{2-\alpha^{2t}}{1-\alpha^{2}}\beta^2\sigma_X^2 + 2\sigma_Y^2 \bigg).
\end{align*}
\end{proof}

\section{Data}\label{sec:data}

\subsection{California's Tobacco Control Program}\label{data:tobacco}

California's tobacco control program~\cite{Abadie2010} uses the annual state-level per-capita cigarette sales panel data from 1970 to 2000. Artificial control units $A_{s, t}$ are created in our simulated study, where $s$ denotes a hypothetical state and $t$ indicates time, and then are added in the donor pool. Then, we perform the standard synthetic control analysis, and check whether \CCMSCM\ or SCM select the artificial units to construct synthetic California. 

We use time series templates to generate artificial control units. In particular, we create 39 artificial states $A_{s, t}$ (the same number of states in original study) with corresponding panel data and four predictors of the outcome variable. The panel data are generated from multiple sets of noisy copies from the template and the predictors are from original tobacco data but with permuted indices for each artificial state. We add $A_{s,t}$ to the original pool to construct a new pool including 77 control units. We also apply moderate data transformations to ensure these adversaries sizable but unrelated to original data, and multiple sets of artificial control units are generated. We run simulations for each set of adversaries and display result distributions with box plots. The template data are described as follows.

\paragraph{Unemployment.} The unemployed percent of US labor force data include annual average employment status of the civilian population from 1976 to 2016, giving 41 years of data for 51 states.~\footnote{Data collected from the Local Area Unemployment Statistics (LAUS) program of the Bureau of Labor Statistics (BLS)~\citep{BLS2018}~\url{https://www.bls.gov/lau/staadata.txt}} We define artificial units as $ A_{s,t} = kU_{s,t}$, where $k$ is a scalar, and $U_{s, t}$ is unemployment for state $s$ at time $t$. To make the data sizable with the tobacco data, $k$ is set to be 6. To generate multiple sets of artificial control units, we select the starting year between 1976 to 1986 and take the following 31 data points as 31 years of data for each state, which gives 11 sets of artificial control units. The simulation results are obtained over 11 runs.

\begin{figure}[!h]
	\centering
	\includegraphics[width=.9\textwidth]{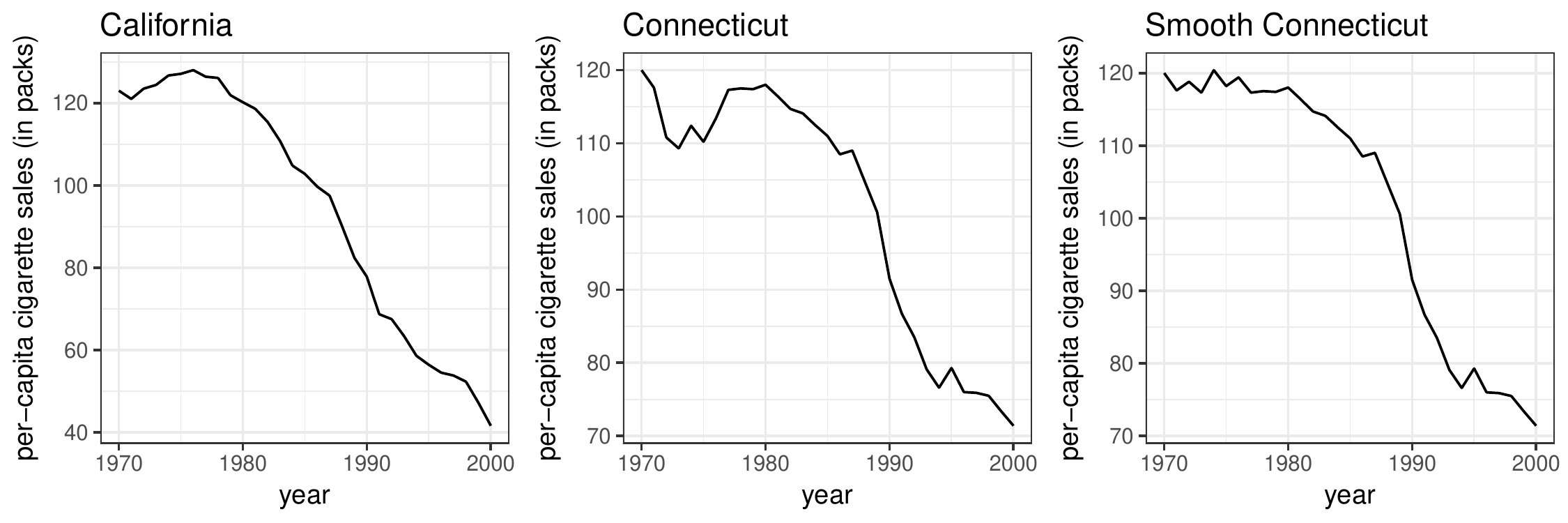}
	\vspace{-0.1in}\caption{Complementary to Figure 3 in the main paper: Per-capita cigarette sales trends for California, Connecticut, and Smooth Connecticut.}
	\label{fig5}  
\end{figure}

\subsection{Brexit Vote}\label{data:brexit}

We picked 30 OECD-member countries as controls, and UK as the treated unit. We collected quarterly real GDP data of these countries from the OECD Economic Outlook database (June 2017) from 1995Q1 to 2018Q4~\footnote{\url{https://stats.oecd.org/index.aspx?DataSetCode=EO}}, where data from 2017Q4 till 2018Q4 are forecasts. The whole quarterly GDP data has 96 data points, and the first 86 points are before Brexit vote. It is assumed that the treatment took form after 2016Q2, and the countries in the donor pool are not affected by the treatment. We also collected predictors of outcome variable such as private consumption, investment, inflation rate, interest rate, and exchange rate.  

We normalized the time series for each country by dividing the time series by its 1995 average and then taking logarithm of that time series to generate the approximately zero starting point in 1995. The predictors of outcome variable include:
\begin{compactenum}[(a)]
	\item Real private consumption: the sum of real final consumption expenditure of both households and non-profit institutions serving households, from 1997Q1 to 2017Q2. %31*82=2542
	\item Real investment: total gross fixed capital formation, from 1995Q1 to 2017Q2. %31*90=2790
	\item Net exports: the external balance of goods and services, from 1997Q1 to 2017Q2. %31*82=2542
	\item Inflation series: the change in the Consumer Price Index (CPI), from 1998Q1 to 2017Q3. %31*79=2449
	\item Quarterly short-term nominal interest rates: quarterly averages of monthly values, from 2002Q1 to 2017Q4. %31*64=1984
	\item Nominal exchange rates: from 1997Q1 to 2018Q4. %only have annual and monthly data. 
\end{compactenum}

The artificial control units are generated in the same way as the example of California's tobacco control program. We use time series template and create 31 artificial countries $A_{s, t}$ (same number of control countries in the original study) with corresponding panel data and six predictors of the outcome variable. We add $A_{s,t}$ to the original pool to construct a new pool including 61 control units. We also apply moderate data transformations to ensure these adversaries sizable and unrelated to original panel data. The details on how to generate the adversaries are described as follows.

\paragraph{Calls.}
We use the calls data collected from the Monthly average daily calls to directory assistance from Jan 1962 to Dec 1976~\footnote{\url{https://datamarket.com/data/set/22yq}} as the template for our adversarial attack. We choose the first 106 data points from this series due to its similar trend with the brexit data, which gives 11 different sets of templates by choosing different starting points. For each template, we fit an autoregressive model and create 31 noisy copies as 31 artificial countries by adding Gaussian noise to them. The simulation results are obtained over 11 runs.

\section{Discussion on CCM}
\label{sec:discuss}

Due to the success of CCM in quantifying dynamical relationships~\citep{Sugihara2012,Deyle6430}, it may be tempting to consider CCM as a method for causal inference. We recommend putting more thoughts before applying this idea. To illustrate why, we apply CCM on causal relationship detection tasks from the benchmark dataset \texttt{CauseEffectPairs}~\citep{Mooij2016}, which contains time series pairs that are known a priori to be causal or not. In practice, time series are normalized before applying CCM on them to ensure all series have the same magnitude for comparison and avoid constructing a distorted state space~\citep{Chang2017}.

\begin{figure}[!h]
	\centering
	\includegraphics[width=.9\textwidth]{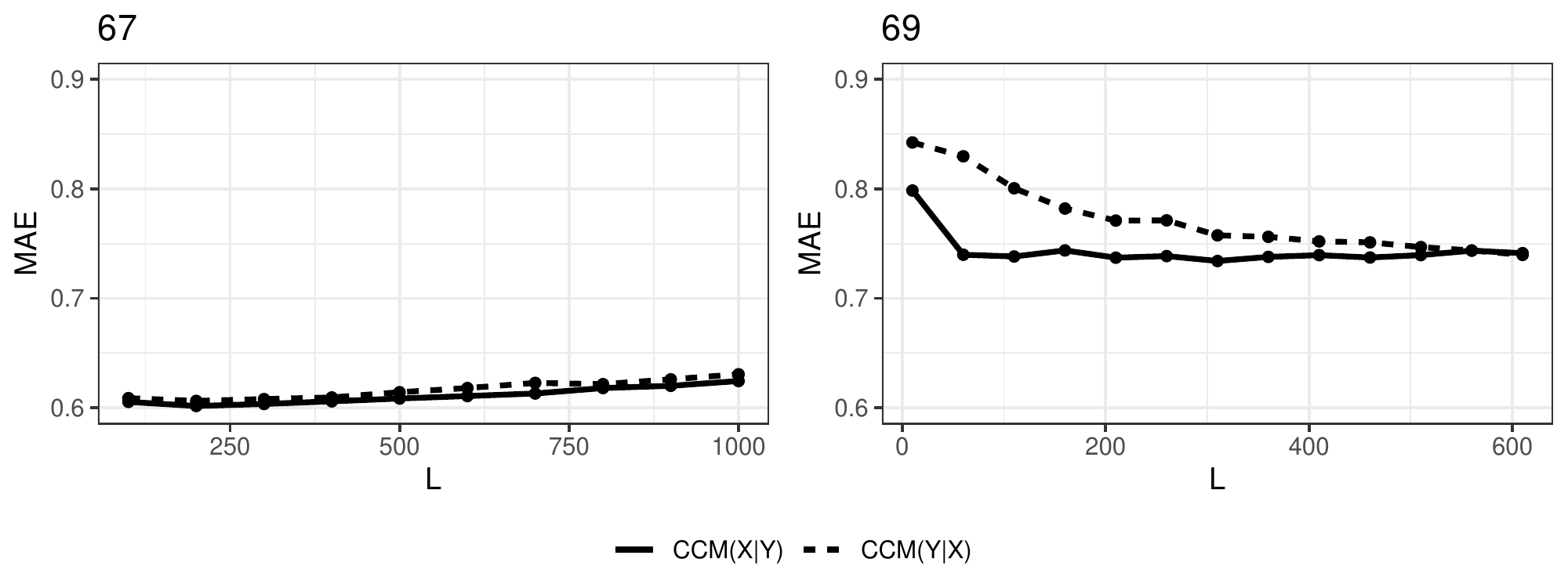}
	\vspace{-0.1in}\caption{CCM results from two pairs. The number in the title of each figure corresponds to the pair index. The ground truths are: 67 ($X\to Y$), 69 ($Y\to X$). }\label{fig: ccm-tue-1}  
\end{figure}

Two cases where CCM fails to detect the true direction of causality are shown in Figure~\ref{fig: ccm-tue-1}. Pair 67 is the financial time series about stock returns from two companies in which one stock is believed to depend on the other. We can see that the CCM score fails to visually converge as library size $L$ increases. By inspection, the time series are close to random walks. Since CCM theory mainly applies on deterministic or chaotic dynamical systems, it is not reliable as a standalone causal inference method in systems dominated by noise.
Another example is Pair 69 in the data of indoor and outdoor temperature. Here, the ground truth is that outdoor temperature variable $Y$ drives the indoor temperature variable $X$, indicating that the dotted curve should converge faster than the solid curve in the right subplot of Figure~\ref{fig: ccm-tue-1}. However, CCM gives the opposite causal direction result. A possible explanation might be that temperature is periodic since it has been suggested that strong periodicity could undermine the effectiveness of CCM~\citep{Chang2017}.

\begin{figure}[!h]
	\centering
	\includegraphics[width=.9\textwidth]{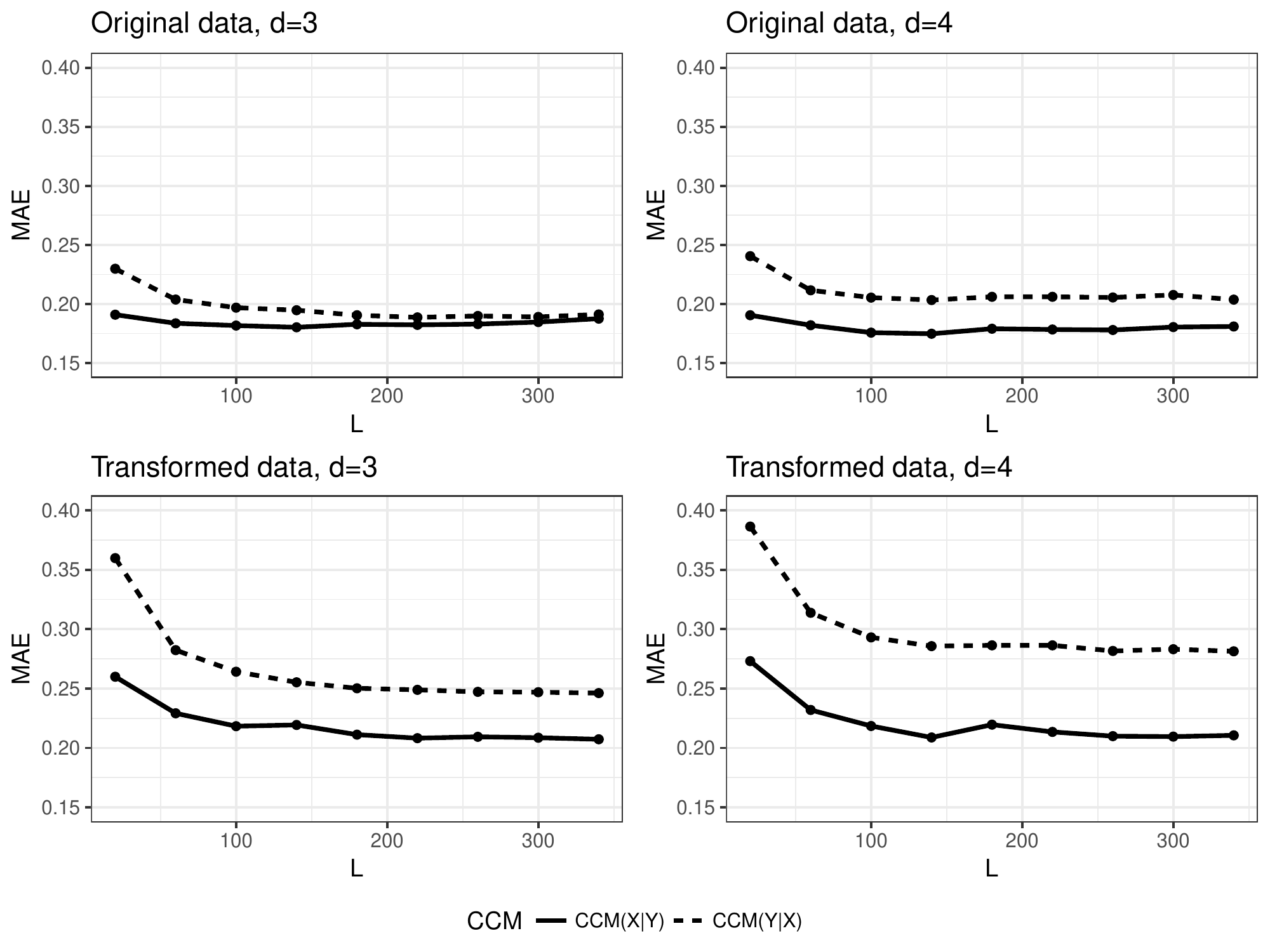}
	\vspace{-0.1in}\caption{CCM results on pair 68. The ground truth is $Y\to X$.}\label{fig:ccm-tue-2}  
\end{figure}

Another practical aspect is that hyperparameters, such as the embedding dimension $d$ and time delay $\tau$, should be carefully chosen. To illustrate this, we consider Pair 68 in the data of internet connections and traffic, where $X$ is bytes sent and $Y$ is number of http connections. Figure~\ref{fig:ccm-tue-2} shows CCM results for this pair with simple data transformations and with varying the embedding dimension $d$. 

Although CCM uncovers the correct causal detection with the original data under embedding dimension $d=3$, the result is not strong enough. Moreover, CCM detects a wrong causal direction when the embedding dimension is set to $d=4$.
We note that the results improve with transformations, say, $\log$ transforms. Optimality of embedding methods and parameter tuning are currently active research areas~\citep{Rosenstein1994,Small2004,Garland2015}.

\bibliography{ccm_syn}
\bibliographystyle{apalike}